\documentclass[longauth]{aa}  

\usepackage{graphicx}
\usepackage{txfonts}
\usepackage{xspace}
\usepackage{color}
\usepackage{url}
\usepackage{hyperref}
\usepackage{comment}
\usepackage{multirow}

\usepackage{newtxtext,newtxmath}
\usepackage{color}

\newcommand{\revision}[1]{{\color{black} #1}}

\usepackage[normalem]{ulem}

\begin{document}

\title{Polarization of reflected X-ray emission from the\\ Sgr~A molecular complex: Multiple flares, multiple sources?}

\author{
Ildar Khabibullin \inst{\ref{in:USM},\ref{in:MPA},\ref{in:IKI}}
\and Eugene Churazov \inst{\ref{in:MPA},\ref{in:IKI}}
\and Riccardo Ferrazzoli \inst{\ref{in:INAF-IAPS}} 
\and Philip Kaaret \inst{\ref{in:NASA-MSFC}} 
\and Jeffery J. Kolodziejczak \inst{\ref{in:NASA-MSFC}} 
\and Fr\'{e}d\'{e}ric~Marin \inst{\ref{in:Strasbourg2}} 
\and Rashid Sunyaev \inst{\ref{in:MPA},\ref{in:IKI}}
\and Jiri Svoboda \inst{\ref{in:CAS-ASU}}
\and Alexey Vikhlinin \inst{\ref{in:CfA},\ref{in:IKI}} 
\and Thibault Barnouin \inst{\ref{in:Strasbourg2}} 
\and Chien-Ting Chen \inst{\ref{in:USRA-MSFC}}
\and Enrico Costa \inst{\ref{in:INAF-IAPS}} 
\and Laura Di Gesu \inst{\ref{in:ASI}} 
\and Alessandro Di Marco \inst{\ref{in:INAF-IAPS}}
\and Steven R. Ehlert \inst{\ref{in:NASA-MSFC}}
\and William Forman \inst{\ref{in:CfA}} 
\and Dawoon E. Kim \inst{\ref{in:INAF-IAPS}}  
\and Ralph Kraft \inst{\ref{in:CfA}} 
\and W. Peter Maksym \inst{\ref{in:NASA-MSFC}}
\and Giorgio Matt  \inst{\ref{in:UniRoma3}}  
\and Juri  Poutanen \inst{\ref{in:Turku}}
\and Paolo Soffitta \inst{\ref{in:INAF-IAPS}} 
\and Douglas  A. Swartz \inst{\ref{in:USRA-MSFC}}
\and Ivan  Agudo \inst{\ref{in:CSIC-IAA}}
\and Lucio Angelo Antonelli \inst{\ref{in:INAF-OAR},\ref{in:ASI-SSDC}}
\and Luca  Baldini \inst{\ref{in:INFN-PI},\ref{in:UniPI}}
\and Wayne H. Baumgartner \inst{\ref{in:NRL}}
\and Ronaldo  Bellazzini \inst{\ref{in:INFN-PI}}
\and Stefano  Bianchi \inst{\ref{in:UniRoma3}}
\and Stephen D. Bongiorno \inst{\ref{in:NASA-MSFC}}
\and Raffaella  Bonino \inst{\ref{in:INFN-TO},\ref{in:UniTO}}
\and Alessandro  Brez \inst{\ref{in:INFN-PI}}
\and Niccolo  Bucciantini \inst{\ref{in:INAF-Arcetri},\ref{in:UniFI},\ref{in:INFN-FI}}
\and Fiamma  Capitanio \inst{\ref{in:INAF-IAPS}}
\and Simone  Castellano \inst{\ref{in:INFN-PI}}
\and Elisabetta  Cavazzuti \inst{\ref{in:ASI}}
\and Stefano  Ciprini \inst{\ref{in:INFN-Roma2},\ref{in:ASI-SSDC}}
\and Alessandra  De Rosa \inst{\ref{in:INAF-IAPS}}
\and Ettore  Del Monte \inst{\ref{in:INAF-IAPS}}
\and Niccolò  Di Lalla \inst{\ref{in:Stanford}}
\and Immacolata  Donnarumma \inst{\ref{in:ASI}}
\and Victor  Doroshenko \inst{\ref{in:Tuebingen}}
\and Michal  Dovciak \inst{\ref{in:CAS-ASU}}
\and Teruaki  Enoto \inst{\ref{in:RIKEN}}
\and Yuri  Evangelista \inst{\ref{in:INAF-IAPS}}
\and Sergio  Fabiani \inst{\ref{in:INAF-IAPS}}
\and Javier A. Garcia \inst{\ref{in:NASA-GSFC}}
\and Shuichi  Gunji \inst{\ref{in:Yamagata}}
\and Kiyoshi  Hayashida \inst{\ref{in:Osaka}}
\and Jeremy  Heyl \inst{\ref{in:UBC}}
\and Wataru  Iwakiri \inst{\ref{in:Chiba}}
\and Svetlana G. Jorstad \inst{\ref{in:BU},\ref{in:SPBU}}
\and Vladimir  Karas \inst{\ref{in:CAS-ASU}}
\and Fabian  Kislat \inst{\ref{in:UNH}}
\and Takao  Kitaguchi \inst{\ref{in:RIKEN}}
\and Henric  Krawczynski \inst{\ref{in:WUStL}}
\and Fabio  La Monaca \inst{\ref{in:INAF-IAPS},\ref{in:UniRoma2}}
\and Luca  Latronico \inst{\ref{in:INFN-TO}}
\and Ioannis  Liodakis \inst{\ref{in:IA-FORTH},\ref{in:NASA-MSFC}}
\and Simone  Maldera \inst{\ref{in:INFN-TO}}
\and Alberto  Manfreda \inst{\ref{in:INFN-NA}}
\and Alan P. Marscher \inst{\ref{in:BU}}
\and Herman L. Marshall \inst{\ref{in:MIT}}
\and Francesco  Massaro \inst{\ref{in:INFN-TO},\ref{in:UniTO}}
\and Ikuyuki  Mitsuishi \inst{\ref{in:Nagoya}}
\and Tsunefumi  Mizuno \inst{\ref{in:Hiroshima}}
\and Fabio  Muleri \inst{\ref{in:INAF-IAPS}}
\and Michela  Negro \inst{\ref{in:LSU}}
\and Chi-Yung  Ng \inst{\ref{in:HKU}}
\and Stephen L. O'Dell \inst{\ref{in:NASA-MSFC}}
\and Nicola  Omodei \inst{\ref{in:Stanford}}
\and Chiara  Oppedisano \inst{\ref{in:INFN-TO}}
\and Alessandro  Papitto \inst{\ref{in:INAF-OAR}}
\and George G. Pavlov \inst{\ref{in:PSU}}
\and Abel Lawrence Peirson \inst{\ref{in:Stanford}}
\and Melissa  Pesce-Rollins \inst{\ref{in:INFN-PI}}
\and Pierre-Olivier  Petrucci \inst{\ref{in:Grenoble}}
\and Maura  Pilia \inst{\ref{in:INAF-OAC}}
\and Andrea  Possenti \inst{\ref{in:INAF-OAC}}
\and Simonetta  Puccetti \inst{\ref{in:ASI-SSDC}}
\and Brian D. Ramsey \inst{\ref{in:NASA-MSFC}}
\and John  Rankin \inst{\ref{in:INAF-OAB}}
\and Ajay  Ratheesh \inst{\ref{in:INAF-IAPS},\ref{in:PRL}}
\and Oliver J. Roberts \inst{\ref{in:USRA-MSFC}}
\and Roger W. Romani \inst{\ref{in:Stanford}}
\and Carmelo  Sgro \inst{\ref{in:INFN-PI}}
\and Patrick  Slane \inst{\ref{in:CfA}}
\and Gloria  Spandre \inst{\ref{in:INFN-PI}}
\and Toru  Tamagawa \inst{\ref{in:RIKEN}}
\and Fabrizio  Tavecchio \inst{\ref{in:INAF-OAB}}
\and Roberto  Taverna \inst{\ref{in:UniPD}}
\and Yuzuru  Tawara \inst{\ref{in:Nagoya}}
\and Allyn F. Tennant \inst{\ref{in:NASA-MSFC}}
\and Nicholas E. Thomas \inst{\ref{in:NASA-MSFC}}
\and Francesco  Tombesi \inst{\ref{in:UniRoma2},\ref{in:INFN-Roma2}}
\and Alessio  Trois \inst{\ref{in:INAF-OAC}}
\and Sergey S. Tsygankov \inst{\ref{in:Turku}}
\and Roberto  Turolla \inst{\ref{in:UniPD},\ref{in:MSSL}}
\and Jacco  Vink \inst{\ref{in:Amsterdam}}
\and Martin C. Weisskopf \inst{\ref{in:NASA-MSFC}}
\and Kinwah  Wu \inst{\ref{in:MSSL}}
\and Fei  Xie \inst{\ref{in:GXU},\ref{in:INAF-IAPS}}
\and Silvia  Zane \inst{\ref{in:MSSL}}
}

\institute{
Universitäts-Sternwarte, Fakultät fuer Physik, Ludwig-Maximilians-Universität Muenchen, Scheinerstr.1, 81679 Muenchen, Germany\label{in:USM}
\and
Max Planck Institute for Astrophysics, Karl-Schwarzschild-Str. 1, D-85741 Garching, Germany \label{in:MPA}
\and
Space Research Institute (IKI), Profsoyuznaya 84/32, Moscow 117997, Russia\label{in:IKI}
\and
INAF Istituto di Astrofisica e Planetologia Spaziali, Via del Fosso del Cavaliere 100, 00133 Roma, Italy \label{in:INAF-IAPS}
\and
NASA Marshall Space Flight Center, Huntsville, AL 35812, USA \label{in:NASA-MSFC}
\and
Universit\'{e} de Strasbourg, CNRS, Observatoire Astronomique de Strasbourg, UMR 7550, 67000 Strasbourg, France \label{in:Strasbourg2}
\and
Astronomical Institute of the Czech Academy of Sciences, Bo\v{c}n\'{i} II 1401/1, 14100 Praha 4, Czech Republic \label{in:CAS-ASU}
\and
Center for Astrophysics, Harvard \& Smithsonian, 60 Garden St, Cambridge, MA 02138, USA \label{in:CfA}
\and
Science and Technology Institute, Universities Space Research Association, Huntsville, AL 35805, USA \label{in:USRA-MSFC}
\and
Agenzia Spaziale Italiana, Via del Politecnico snc, 00133 Roma, Italy \label{in:ASI}
\and
Dipartimento di Fisica, Università degli Studi di Roma "La Sapienza", Piazzale Aldo Moro 5, 00185 Roma, Italy \label{in:LaSap}
\and
Istituto Nazionale di Fisica Nucleare, Sezione di Roma ``Tor Vergata'', Via della Ricerca Scientifica 1, 00133 Roma, Italy \label{in:INFN-Roma2}
\and
Dipartimento di Matematica e Fisica, Universit\`a degli Studi Roma Tre, via della Vasca Navale 84, 00146 Roma, Italy \label{in:UniRoma3}
\and
Department of Physics and Astronomy, 20014 University of Turku, Finland \label{in:Turku}
\and
Instituto de Astrof\'{i}sica de Andaluc\'{i}a -- CSIC, Glorieta de la Astronom\'{i}a s/n, 18008 Granada, Spain \label{in:CSIC-IAA}
\and
INAF Osservatorio Astronomico di Roma, Via Frascati 33, 00040 Monte Porzio Catone (RM), Italy \label{in:INAF-OAR}
\and
Space Science Data Center, Agenzia Spaziale Italiana, Via del Politecnico snc, 00133 Roma, Italy \label{in:ASI-SSDC}
\and
Istituto Nazionale di Fisica Nucleare, Sezione di Pisa, Largo B. Pontecorvo 3, 56127 Pisa, Italy \label{in:INFN-PI}
\and
Dipartimento di Fisica, Universit\`{a} di Pisa, Largo B. Pontecorvo 3, 56127 Pisa, Italy \label{in:UniPI}
\and
Naval Research Laboratory, 4555 Overlook Ave. SW, Washington, DC 20375, USA \label{in:NRL}
\and
Istituto Nazionale di Fisica Nucleare, Sezione di Torino, Via Pietro Giuria 1, 10125 Torino, Italy \label{in:INFN-TO}
\and
Dipartimento di Fisica, Universit\`{a} degli Studi di Torino, Via Pietro Giuria 1, 10125 Torino, Italy \label{in:UniTO}
\and
INAF Osservatorio Astrofisico di Arcetri, Largo Enrico Fermi 5, 50125 Firenze, Italy \label{in:INAF-Arcetri}
\and
Dipartimento di Fisica e Astronomia, Universit\`{a} degli Studi di Firenze, Via Sansone 1, 50019 Sesto Fiorentino (FI), Italy \label{in:UniFI}
\and
Istituto Nazionale di Fisica Nucleare, Sezione di Firenze, Via Sansone 1, 50019 Sesto Fiorentino (FI), Italy \label{in:INFN-FI}
\and
Department of Physics and Kavli Institute for Particle Astrophysics and Cosmology, Stanford University, Stanford, California 94305, USA \label{in:Stanford}
\and
Institut fuer Astronomie und Astrophysik, Universität Tuebingen, Sand 1, 72076 Tuebingen, Germany \label{in:Tuebingen}
\and
RIKEN Cluster for Pioneering Research, 2-1 Hirosawa, Wako, Saitama 351-0198, Japan \label{in:RIKEN}
\and
NASA Goddard Space Flight Center, Greenbelt, MD 20771, USA \label{in:NASA-GSFC}
\and
Yamagata University,1-4-12 Kojirakawa-machi, Yamagata-shi 990-8560, Japan \label{in:Yamagata}
\and
Osaka University, 1-1 Yamadaoka, Suita, Osaka 565-0871, Japan \label{in:Osaka}
\and
University of British Columbia, Vancouver, BC V6T 1Z4, Canada \label{in:UBC}
\and
International Center for Hadron Astrophysics, Chiba University, Chiba 263-8522, Japan \label{in:Chiba}
\and
Institute for Astrophysical Research, Boston University, 725 Commonwealth Avenue, Boston, MA 02215, USA \label{in:BU}
\and
Department of Astrophysics, St. Petersburg State University, Universitetsky pr. 28, Petrodvoretz, 198504 St. Petersburg, Russia \label{in:SPBU}
\and
Department of Physics and Astronomy and Space Science Center, University of New Hampshire, Durham, NH 03824, USA \label{in:UNH}
\and
Physics Department and McDonnell Center for the Space Sciences, Washington University in St. Louis, St. Louis, MO 63130, USA \label{in:WUStL}
\and
Dipartimento di Fisica, Universit\`{a} degli Studi di Roma ``Tor Vergata'', Via della Ricerca Scientifica 1, 00133 Roma, Italy \label{in:UniRoma2}
\and
Dipartimento di Fisica, Università degli Studi di Roma “La Sapienza”, Piazzale Aldo Moro 5, 00185 Roma, Italy \label{in:UniRoma1}
\and
Institute of Astrophysics, Foundation for Research and Technology - Hellas, Voutes, 7110, Heraklion, Greece\label{in:IA-FORTH}
\and
Istituto Nazionale di Fisica Nucleare, Sezione di Napoli, Strada Comunale Cinthia, 80126 Napoli, Italy \label{in:INFN-NA}
\and
MIT Kavli Institute for Astrophysics and Space Research, Massachusetts Institute of Technology, 77 Massachusetts Avenue, Cambridge, MA 02139, USA \label{in:MIT}
\and
Graduate School of Science, Division of Particle and Astrophysical Science, Nagoya University, Furo-cho, Chikusa-ku, Nagoya, Aichi 464-8602, Japan \label{in:Nagoya}
\and
Hiroshima Astrophysical Science Center, Hiroshima University, 1-3-1 Kagamiyama, Higashi-Hiroshima, Hiroshima 739-8526, Japan \label{in:Hiroshima}
\and
Department of Physics and Astronomy, Louisiana State University, Baton Rouge, LA 70803, USA \label{in:LSU}
\and
Department of Physics, University of Hong Kong, Pokfulam, Hong Kong \label{in:HKU}
\and
Department of Astronomy and Astrophysics, Pennsylvania State University, University Park, PA 16801, USA \label{in:PSU}
\and
Universit\'{e} Grenoble Alpes, CNRS, IPAG, 38000 Grenoble, France \label{in:Grenoble}
\and
INAF Osservatorio Astronomico di Cagliari, Via della Scienza 5, 09047 Selargius (CA), Italy \label{in:INAF-OAC}
\and
INAF Osservatorio Astronomico di Brera, via E. Bianchi 46, 23807 Merate (LC), Italy \label{in:INAF-OAB}
\and
Physical Research Laboratory, Thaltej, Ahmedabad, Gujarat 380009, India\label{in:PRL}
\and
Dipartimento di Fisica e Astronomia, Universit\`{a} degli Studi di Padova, Via Marzolo 8, 35131 Padova, Italy \label{in:UniPD}
\and
Mullard Space Science Laboratory, University College London, Holmbury St Mary, Dorking, Surrey RH5 6NT, UK \label{in:MSSL}
\and
Anton Pannekoek Institute for Astronomy \& GRAPPA, University of Amsterdam, Science Park 904, 1098 XH Amsterdam, The Netherlands \label{in:Amsterdam}
\and
Guangxi Key Laboratory for Relativistic Astrophysics, School of Physical Science and Technology, Guangxi University, Nanning 530004, China \label{in:GXU}
}


\titlerunning{Polarization of X-ray echo}
\authorrunning{Khabibullin et al.}
\date{2025}

\abstract{
~~~~~The extended X-ray emission observed in the direction of several molecular clouds in the central molecular zone (CMZ) of our Galaxy exhibits spectral and temporal properties consistent with the X-ray echo scenario. 
This concept postulates that the observed signal is a light-travel-time delayed reflection of a short ($\delta t<$1.5 yr) and bright ($L_{\rm X}>10^{39}~{\rm erg~s^{-1}}$) flare that was most probably produced a few hundred years ago by Sgr~A*. 
This scenario also predicts a distinct polarization signature for the reflected X-ray continuum, with the polarization vector being perpendicular to the direction toward the primary source and the polarization degree being determined by the scattering angle. We report the results of two deep observations of the currently brightest (in reflected emission) molecular complex Sgr~A taken with the Imaging X-ray Polarimetry Explorer (IXPE) in 2022 and 2023. 
We confirm the previous polarization measurement for a large region encompassing the Sgr~A complex with high significance. We also reveal an inconsistent polarization pattern for the brightest reflection region in its center. Specifically, the X-ray polarization from this region is almost perpendicular to the expected direction in the case of Sgr~A* illumination, and it shows a smaller degree of polarization compared to the large region. Taken at face value, this could indicate the simultaneous propagation of several illumination fronts throughout the CMZ, with the origin of one of them not being Sgr~A*. The primary source could be associated with the Arches stellar cluster or a currently unknown source located closer to the illuminated cloud, potentially lowering the required luminosity of the primary source. Although significantly deeper observations with IXPE would be required to unequivocally distinguish between the scenarios, a combination of high-resolution imaging and micro-calorimetric spectroscopy offers an additional promising path forward.  
}




\maketitle

\section{Introduction}
\label{s:introduction}
 X-ray reflection off molecular clouds in the central molecular zone (CMZ) of our Galaxy offers a unique window into the past activity record of the currently dormant supermassive black hole Sgr~A* \citep[][]{1980SvAL....6..353V,1993ApJ...407..606S,1993Natur.364...40M,1996PASJ...48..249K,2013ASSP...34..331P}. The light-travel-delay due to propagation from the primary source to the reflector implies that the currently observed reflected X-ray emission is connected to the primary's source luminosity from decades to centuries ago \citep[as first realized and described for nova eruptions by][]{1901AN....157..201K,1939AnAp....2..271C}.

The apparent variability of the observed diffuse emission {\citep[][]{2004A&A...425L..49R,2007ApJ...656L..69M,2009PASJ...61S.241I,2010ApJ...714..732P,2012A&A...545A..35C,2013A&A...558A..32C,2013PASJ...65...33R,2015ApJ...815..132Z,2017MNRAS.468.2822K,2018A&A...610A..34C,2018A&A...612A.102T,2022MNRAS.509.1605K,2022MNRAS.509.6068K,2025AJ....169..213A,2025ApJ...982L..20B, 2025A&A...695A..52S}} can be best explained by the short flare scenario. It posits that the duration of the primary energetic event responsible for the clouds' illumination cannot exceed the observed duration of reflected emission from the smallest sub-structures, the light-crossing time\footnote{The "scanning" speed of the illumination front propagation along the line of sight varies from $0.5c$ at the back side of the illumination paraboloid to infinity in its front side ($c$ is the speed of light).} of which is relatively short \citep[e.g.,][]{2017MNRAS.468..165C}. 

Since the observed surface brightness of the reflected emission is determined primarily by the distance between the cloud and the source, i.e., the mean density of the molecular gas and flare's fluence (flux integrated over duration of a flare), one can use density estimates obtained from molecular line emission to obtain a lower limit on the average luminosity of the primary source (corresponding to the upper limit on the flare's duration for a fixed value of fluence). Such estimates give us a limit that still allows a flare from an X-ray binary at the Eddington level, $\mathrm{\sim10^{39}~erg~s^{-1}}$, lasting for one year as a possible origin for the X-ray echo. On the other hand, for an Eddington-level flare of Sgr~A*, $\mathrm{\sim10^{44}~erg~s^{-1}}$, the duration of merely an hour would suffice to produce the required X-ray fluence. Although large amplitude \revision{\citep[up to a factor of more than one hundred relative to the quiescent level; see][]{2012ApJ...759...95N,2019ApJ...886...96H}} flares are regularly observed from Sgr~A* in infrared and X-ray monitoring campaigns {\citep[][]{2004A&A...427....1E,2009ApJ...698..676D,2017MNRAS.468.2447P}}, the total energy required to account for the X-ray echo is larger by many orders of magnitude \revision{($\sim10^{47}$ erg versus $\sim10^{39}$ erg)}, and it is unclear if it could have been produced by the same mechanism \citep[e.g.,][]{2013ASSP...34..331P,2017MNRAS.465...45C}.

An important and instrumental prediction of the X-ray reflection scenario is linear polarization of the continuum part of the emission, with the polarization degree determined (in the simplest case of single scattering and unpolarized primary emission) by the cosine of the scattering angle and the polarization angle (PA) being perpendicular to the direction toward the primary emission source {\citep{2002MNRAS.330..817C,2017MNRAS.468..165C,2014MNRAS.441.3170M,2015A&A...576A..19M,2020MNRAS.495.1414K,2020A&A...643A..52D,2021A&A...655A.108F}}. Thus, X-ray polarimetry has long been recognized as a potential way to pin down the location of the primary source and the relative 3D position of the reflecting cloud and help break degeneracies in parameters of the spectral reflection model \citep[e.g., relative abundance of iron responsible for production of the brightest fluorescent line at 6.4 keV,][]{2017MNRAS.471.3293C}. 

\revision{In a more complex scenario, when the X-ray reflection throughout the CMZ is produced not by a single flare but by a set of flares separated in time, or even by multiple independent sources, the spatial pattern of the X-ray reflection signal offers the cleanest way to build the complete illumination picture \citep[e.g.,][]{2014MNRAS.441.3170M,2015A&A...576A..19M}. Finally, intrinsic linear polarization of the primary flare emission might have characteristic imprints on the spatial polarization properties even in the simplest case of a single flare illumination \citep[e.g.,][]{2020MNRAS.498.4379K}. Given that different scenarios predict a variety of temporal, spectral, and polarization relations \citep[driven by dependence of the continuum albedo on the scattering angle and independence of the fluorescent line emission of the geometrical configuration; ][]{2002MNRAS.330..817C}, it is crucial to obtain a sensitive spectrally resolved 
polarization map rather than a single image-integrated quantity.}

The first polarimeteric observation of the currently brightest (in reflected emission) molecular complex Sgr~A \citep[][]{2022MNRAS.509.6068K,2025ApJ...982L..20B,2025A&A...695A..52S} was performed by the Imaging X-ray Polarimetry Explorer \citep[IXPE;][]{2022JATIS...8b6002W} observatory in 2022, and 
it broadly confirmed the predicted scenario of a flare from Sgr~A* some 200 years ago \citep{2023Natur.619...41M}. Even for this brightest complex and rather long effective exposure time of the observation, the level of statistical and systematic noise enabled a polarization measurement 
only for the whole complex and detection of the polarization at a 3$\sigma$ level. 

In order to improve the statistical significance of this result and enable measurements for individual substructures, a second observation was performed in 2023. This doubled the total clean exposure time (after filtering time intervals affected by enhanced background level) at the source position \citep[][]{2024A&A...686A..14C}.

We report the results of the two epochs of deep observations of Sgr~A taken with IXPE. We confirm the previous polarization measurement for a large region encompassing the Sgr~A complex 
with high significance, but we also reveal a possibly inconsistent polarization pattern for the brightest reflection region in its center. Specifically, X-ray polarization from this bright region is almost perpendicular to the expected direction in the case of Sgr~A* illumination, and it shows a smaller degree of polarization compared to the large region.

The paper is organized as follows: We outline the short flare scenario and expected polarization signal in Sect.~\ref{s:shortflare}. The  \textit{Chandra} and IXPE data are described in  Sect.~\ref{s:data}. In Sect.~\ref{s:imaging}, we demonstrate the consistency of the observed emission from the Sgr~A region with the X-ray reflection scenario using epoch-to-epoch variations of the diffuse emission. We present spatially resolved spectro-polarimetric measurements from the combined IXPE dataset in Sect.~\ref{s:polarimetry}. We discuss the results in Sect.~\ref{s:discussion}, and we present the summary and outlook  in Sect.~\ref{s:conclusion}.

\section{Short flare scenario}
\label{s:shortflare}
\begin{figure}
\centering
\includegraphics[width=1.0\columnwidth, trim=3cm 6cm 1cm 3cm]{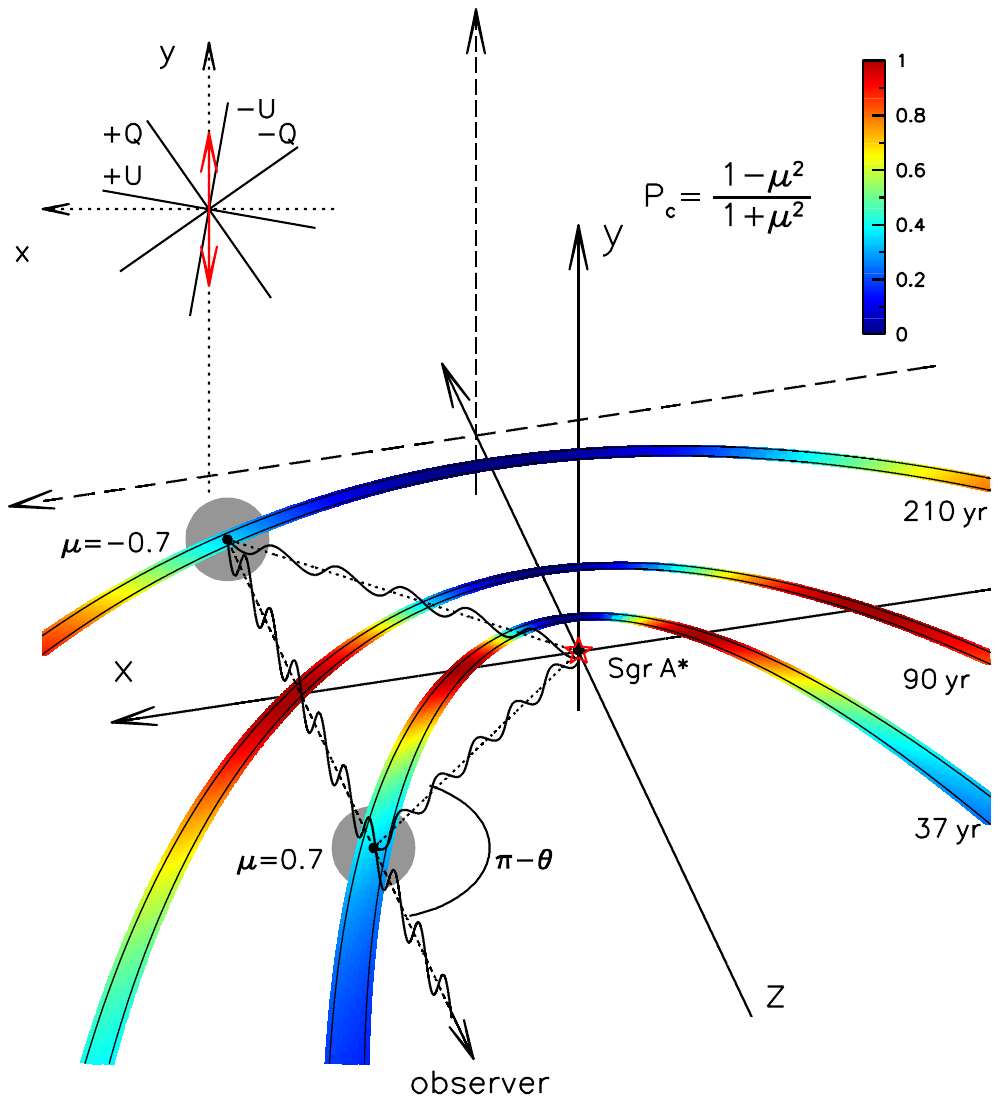}
\caption{Sketch illustrating the polarization of the X-ray emission arising due to scattering of an intrinsically unpolarized short flare. {The coordinate system is chosen so that the primary source, Sgr~A*, is located at the origin, $z$-axis points from the observer toward it, and the $xz$-plane contains the center of the reflecting molecular complex (gray sphere)}. The line of sight and image plane positions of the cloud are connected by the parabolic relation \revision{(Eq.~\ref{eq:parabola}).}
For an intrinsically unpolarized primary source, the electric field vector of the reflected emission is perpendicular to the scattering (i.e., $xz$) plane, and the polarization degree (color coded) is set by the cosine of the scattering angle $\theta$. 
The inset shows the definition of the Stokes parameters used throughout the paper {(with $+Q$ and $+U$ pointing toward the celestial north and northeast, respectively)} and the predicted \revision{direction of electric field vector (red arrow)}. 
}
\label{fig:sketch_pol}
\end{figure}

Let us consider a compact molecular complex that is reflecting X-ray emission from Sgr~A*'s flare that occurred $t_{\rm age}\sim100$ yr ago. Except for the photoionization followed by fluorescent line emission (most importantly the fluorescent line of the neutral iron at 6.4 keV), the key process is Compton scattering, which, in the majority of the relevant cases, might be treated in the optically thin limit \citep[e.g.,][]{2017MNRAS.471.3293C}. The polarization of the reflected X-ray continuum is fully determined by the geometry of the scattering, set by the relative disposition of the cloud and the primary source, if the primary radiation was unpolarized.

We use the coordinate system such that the primary source is located at the origin, the $z$-axis is pointing from the observer toward it, and the reflecting cloud resides in the $xz$-plane (see Fig.~\ref{fig:sketch_pol}). This plane is referred to as the scattering plane in what follows, and the $y$-axis represents the direction normal to it.

The line of sight and image plane coordinates of the reflecting gas are connected by the relation 
\begin{equation}
    Z=\frac{c}{2}t_{\rm age}-\frac{X^2}{2ct_{\rm age}},
    \label{eq:parabola}
\end{equation}
which can be formulated in the dimensionless (and time-independent) form after dividing by the characteristic length scale $R_0=ct_{\rm age}\approx31$pc $(t_{\rm age}/100\,{\rm yr})$ so that $(x,y,z)=(X,Y,Z)/ct_{\rm age}$:
\begin{equation}
    {z}=\frac{1}{2}\left(1-x^2\right).
    \label{eq:paraboloid}
\end{equation}
The distance from the primary source to the cloud equals
\begin{equation}
    R/R_0=\sqrt{x^2+z^2}=\frac{1}{2}\left(1+x^2\right),
    \label{eq:r}
\end{equation}
and the cosine of the scattering angle $\theta$ (see Fig.~\ref{fig:sketch_pol}) is calculated as
\begin{equation}
\mu=\cos\theta=-Z/R=-\left(1-{x}^2\right)/\left(1+{x}^2\right).  
\label{eq:mu}
\end{equation}

For the flare duration $\Delta t\sim 1$ yr $\ll t_{\rm age}$, the thickness of the gas layer contributing to the reflected emission for any line-of-sight direction $(X,Y)$ is small, 
\begin{equation}
    \Delta \mu\approx\frac{d\mu}{dt}\Delta t\sim \Delta t/t_{\rm age}\ll 1,
\end{equation}
so one can neglect variations in the scattering geometry across this layer. The degree of polarization, $P$, depends on the cosine of the scattering angle, $\mu$, 
{(in the limit of Thomson scattering) as
\begin{eqnarray}
P=\frac{1-\mu^2}{1+\mu^2}.
\label{eq:p}
\end{eqnarray}

Thus, knowing the degree of polarization, $P$, one can determine the 3D position of an illuminated cloud. For instance, {$P=31(\pm11)\%$ as measured for the Sgr~A complex by \citealt{2023Natur.619...41M} corresponds to $\theta=137^\circ$, and for $X=25\,{\rm pc}$, one gets $Z\approx 26 \,{\rm pc}$.} That is, the illuminated cloud is farther away from us than Sgr~A$^*$ by this distance. 

{\revision{Using Eq.~\eqref{eq:mu} and Eq.~\eqref{eq:p}}, one can explicitly relate \revision{the age of the flare $t_{\rm age}$} and the measured degree of polarization $P$ as
\begin{eqnarray}
t_{\rm age}=\frac{X}{c}\times \left [ \left ( \frac{1+P}{2P} \right )^{1/2} \pm \left ( \frac{1-P}{2P} \right )^{1/2}\right].
\end{eqnarray}
We used this expression to estimate the age of the flare.

\revision{Given that the degree of polarization depends on the scattering angle, the finite extent $\delta Z\approx\Delta t\times \varv_{z}$ of the illuminated medium along the line of sight might be important for a long flare ($\varv_z=dZ/dt=\frac{c}{2}(1+X^2/R_0^2)$ from Eq.~\eqref{eq:parabola}). This extent can be neglected as long as $\delta Z$ is much smaller than the distance $R$ between the primary source and the scattering cloud. Given that $\varv_z=dZ/dt$ varies between $0.5c$ (scattering cloud behind the source with $X/R_0\approx0$) and $\infty$ (scattering cloud in front of the source with $X/R_0\gg1$), the reflected emission from clouds closer to us than the primary source (in the single-scattering approximation) is short-lived, while the clouds farther away will remain visible for a much longer time. Specializing for the latter case, the condition $\Delta t\ll t_{\rm age}$ is sufficient to neglect the finite extent of the cloud. In the limit of a very long flare illuminating an infinite homogeneous medium (i.e., all scattering angles are present for a given line of sight direction), the resulting polarization is $\frac{1}{3}$ (assuming dipole phase function $\sigma_{\rm sc}\propto1+\cos^2\theta$ for continuum scattering).\footnote{\revision{This can be calculated taking into account that total scattered luminosity per unit length $dL_{\rm sc}/dz\propto \sigma_{\rm sc}/R^2\propto (1+2z^2)/(1+z^2)^2$, while the polarized one as $PdL_{\rm sc}/dz\propto 1/(1+z^2)^2$ from Eq.~\eqref{eq:r},~\eqref{eq:mu} and \eqref{eq:p}, and the corresponding integrals are easily evaluated via $z=\tan \xi$ substitution.}}}

An extension of this basic scenario, allowing for possible intrinsic polarization of the original flare's emission, was considered by \citet{2002MNRAS.330..817C}, who presented explicit expressions for the polarization degree of the scattered emission as a function of the cloud's relative position {\citep[see also][where scattering of anisotropic and polarized time-variable radiation has been modeled]{1987SvAL...13..233G}}. This analysis was expanded by \citet{2020MNRAS.495.1414K} to predict the full polarization pattern of the scattered emission, i.e., the maps of the polarization degree and polarization plane orientation for various configurations of the primary source polarization. In contrast to the case of the unpolarized primary emission, the predicted patterns demonstrate substantial morphological complexity, including non-axisymmetric variations in the intensity, polarization degree, and polarization plane orientation. For instance, in this case, the electric field vector is not necessarily perpendicular to the line connecting the primary source and the reflecting cloud. Also, given that, contrary to the scattered continuum emission, fluorescent line emission is insensitive to the intrinsic polarization of the initial flare, correlated spatial variations in polarization and spectral properties (e.g., equivalent width of the fluorescent line) are expected in this case \citep{2020MNRAS.495.1414K}.

\section{Data}
\label{s:data}

\subsection{IXPE data}
In this work, we used the data of two IXPE observations of the Galactic Center region, more specifically, a complex of molecular clouds $\sim 0.1$ degrees to the east of Sgr~A* \citep{2023Natur.619...41M} -- the Sgr~A complex
\revision{centered on (RA,~Dec)=(266.566$^\circ,-28.891^\circ$), i.e., $(l,b)=(0.112^\circ,-0.097^\circ)$ in Galactic coordinates}. These observations with a total clean (i.e., after excluding periods of enhanced particle background and solar flares) exposure time of $1.8$~Msec, were performed in two parts --  in February 2022 and September 2023. These two observations were targeted at (RA, Dec)=(266.51$^\circ,-28.89^\circ$) and (266.57$^\circ,-28.89^\circ$), respectively, with an intentional offset between their aim-points of $\sim 3'$.  

The raw IXPE data were processed with the standard pipeline { (version 1.9.2 with \texttt{HEASOFT} v6.34)\footnote{\url{https://heasarc.gsfc.nasa.gov/docs/software/heasoft/}} and CALDB version of May 2025}. The output FITS files of this pipeline contain the event-by-event Stokes parameters \citep[see][]{2015APh....68...45K} from which the polarization observables of the X-ray radiation can be derived. The data products are publicly available for use by the international astrophysics community at the High-Energy Astrophysics Science Archive Research Center\footnote{\url{https://heasarc.gsfc.nasa.gov/docs/ixpe/archive/}} (HEASARC, at the NASA Goddard Space Flight Center). An energy-dependent particle background rejection algorithm {(\texttt{IXPE-Background}\footnote{\url{https://github.com/aledimarco/IXPE-background}} tool, v2.3)}
was applied to these level-2 event files; it allows the removal of $\sim40\%$ of the instrumental background \citep{2023AJ....165..143D}. Observation periods affected by increased background due to solar activity were removed (see \citealt{2023Natur.619...41M} for additional details on the procedure). The second epoch observation was performed with a smaller dithering amplitude, aiming at achieving a higher sensitivity in the central part of the field of view (FoV). {As a result, the brightest compact (i.e., spatially unresolved) source, the Arches cluster, which was used to adjust the astrometry of the first observation, was not visible in the nominal FoV. The Arches cluster was, however, detectable near a corner of the FoV of Detector Unit 1 (DU1) in the Level 1 (L1) event data. Custom analysis of L1 data to determine attitude offsets for the individual observation segments yielded a means for consistent astrometric correction. These attitude corrections were applied after the standard pipeline processing to generate the final Level~2 products, used here and available in the HEASARC.} The resulting astrometry is fully sufficient for singling out the emission of the brightest reflection region, as well as {pulsar wind nebula (PWN)} G0.13-0.11. { No additional weighting or binning of the data was included in the spatial (broad-band) or spectropolarimetric analyses. Long IXPE observations of the Circinus galaxy were used to model quiescent background spectrum, i.e., the sum of the sky and detector backgrounds in the area not contaminated by the target source (after applying identical filtering routines), and subtract it in the spectral analysis of the Stokes $I$ data (consistently with the procedures performed and described in \citealt[][]{2023Natur.619...41M}).}

\subsection{Chandra data}

High spatial resolution X-ray images were obtained with the \textit{Chandra} Observatory \citep[][]{2000SPIE.4012....2W} over multiple observing campaigns from 2000 to 2024.\revision{The total exposure of the \textit{Chandra} pointings having Sgr~A complex located in the central region of the FoV is $\approx$1.5~Msec.} The \textit{Chandra} data reduction follows a standard procedure based on the latest versions of the data reduction software (CIAO v.\ 4.14) and calibration (CALDB v.\ 4.9.8). Our particular approach and analysis steps are described in detail in \cite{2009ApJ...692.1033V}. Briefly, they include the identification and removal of high background periods, the correction of photon energies for the time and detector temperature dependence of the charge transfer inefficiency and gain, \revision{and the creation of matching background datasets using blank sky observations with exposure times similar to the GC pointings. For the analysis presented here, we used the combined flat-fielded and background-subtracted \textit{Chandra} images in the 3--8~keV band and spectra extracted in several regions of interest.} Following the standard approach for analyzing \textit{Chandra} spectra of extended sources, we have generated the spectral response files that combine the position-dependent ACIS calibration with the weights proportional to the observed brightness.

\section{X-ray images, light curves, and spectra: Epoch-to-epoch variations from \textit{Chandra}} 
\label{s:imaging}

\revision{The reflection scenario for the Sgr~A complex is well established and tested on the data of extensive \textit{Chandra} and \textit{XMM-Newton} monitoring campaigns \citep[][]{2010ApJ...714..732P, 2013ASSP...34..331P,2013A&A...558A..32C,2017MNRAS.465...45C,2018A&A...612A.102T,2025AJ....169..213A}.} Here, we are interested in the quantitative characterization of the reflection signal from several selected regions in order to use this as a basis for the spectropolarimetric analysis of the IXPE data. 

\revision{Due to the expected spectral shape of the reflected emission \citep[][]{1996AstL...22..648S,2017MNRAS.468..165C} and high foreground interstellar absorption toward the Galactic Center, the optimal band is commonly selected as  3--8 keV. This band contains both the reflection continuum and the most prominent fluorescent line of neutral (or only weakly ionized) iron at 6.4 keV. Also, it provides good leverage for spectral decomposition analysis needed to single out the `hot plasma' ($kT\sim7$ keV) component \citep[the bulk of which is produced by the unresolved population of accreting white dwarfs, so called Galactic ridge and bulge emission,][]{2009Natur.458.1142R}, as well an intermediate temperature component ($kT\sim$2--3 keV), which is especially bright in the direction of the Sgr~A complex. Finally, this band is almost insensitive to the dramatic decrease in Chandra's response efficiency at low energies due to the build-up of the absorbing contaminant layer, allowing one a simple comparison between the epochs just at the band-integrated maps level \citep[e.g.,][]{2017MNRAS.465...45C}.}
\subsection{Imaging}
\begin{figure*}
\centering
\includegraphics[angle=0,trim=1cm 8.5cm 1cm 6.5cm,clip,width=2.\columnwidth]{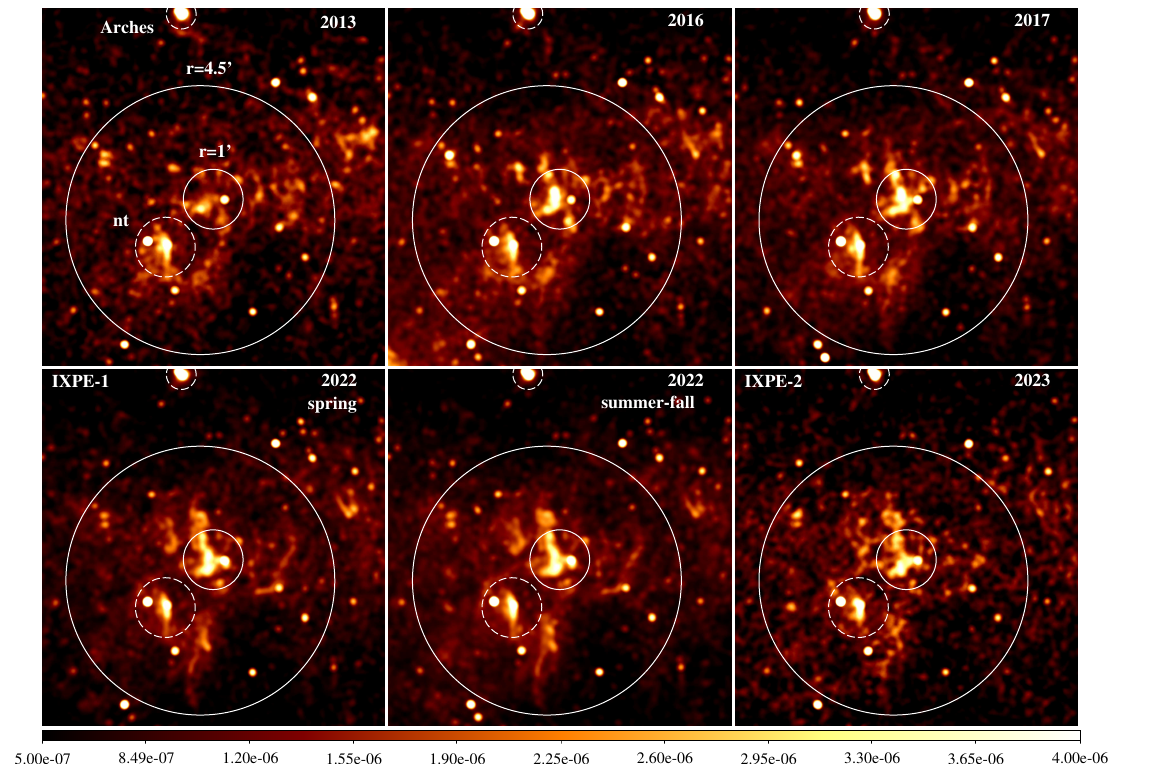}
\caption{\textit{Chandra} images of the Sgr~A complex in the 3--8 keV band taken at different epochs starting from 2013 to 2023. The images have been \revision{smoothed with a $\sigma=5''$ Gaussian to highlight the morphology of the diffuse emission} (${\rm ct~s^{-1}~pixel^{-1}}$, linear scale, Galactic coordinates). The epochs marked as IXPE-1 and IXPE-2 most closely correspond to the two epochs of IXPE observations in 2022 and 2023. The large circle has a 4.5$\arcmin$ radius and was used in Paper~I for the extraction of the polarization signal after masking the region around the pulsar and related nonthermal nebula G0.13-0.11 (marked with { "nt" and a dashed circle} of 1$\arcmin$ radius). The brightest region of the Sgr~A complex is marked with a small solid circle ("$r=1\arcmin$," also 1$\arcmin$ radius). The location of the Arches cluster is also marked in the top part of the image. Clear time variations of the diffuse emission are visible, indicating regions likely dominated by the reflected emission  \revision{(the data from 2013 and 2023 have lower exposure, resulting in enhanced photon-counting noise fluctuations).}
Sgr~A* lies outside the region shown, at a few arc minutes from the right boundary. 
}
\label{fig:chandra_epochs}
\end{figure*}
\begin{figure}
\centering
\includegraphics[angle=0,width=1.\columnwidth]{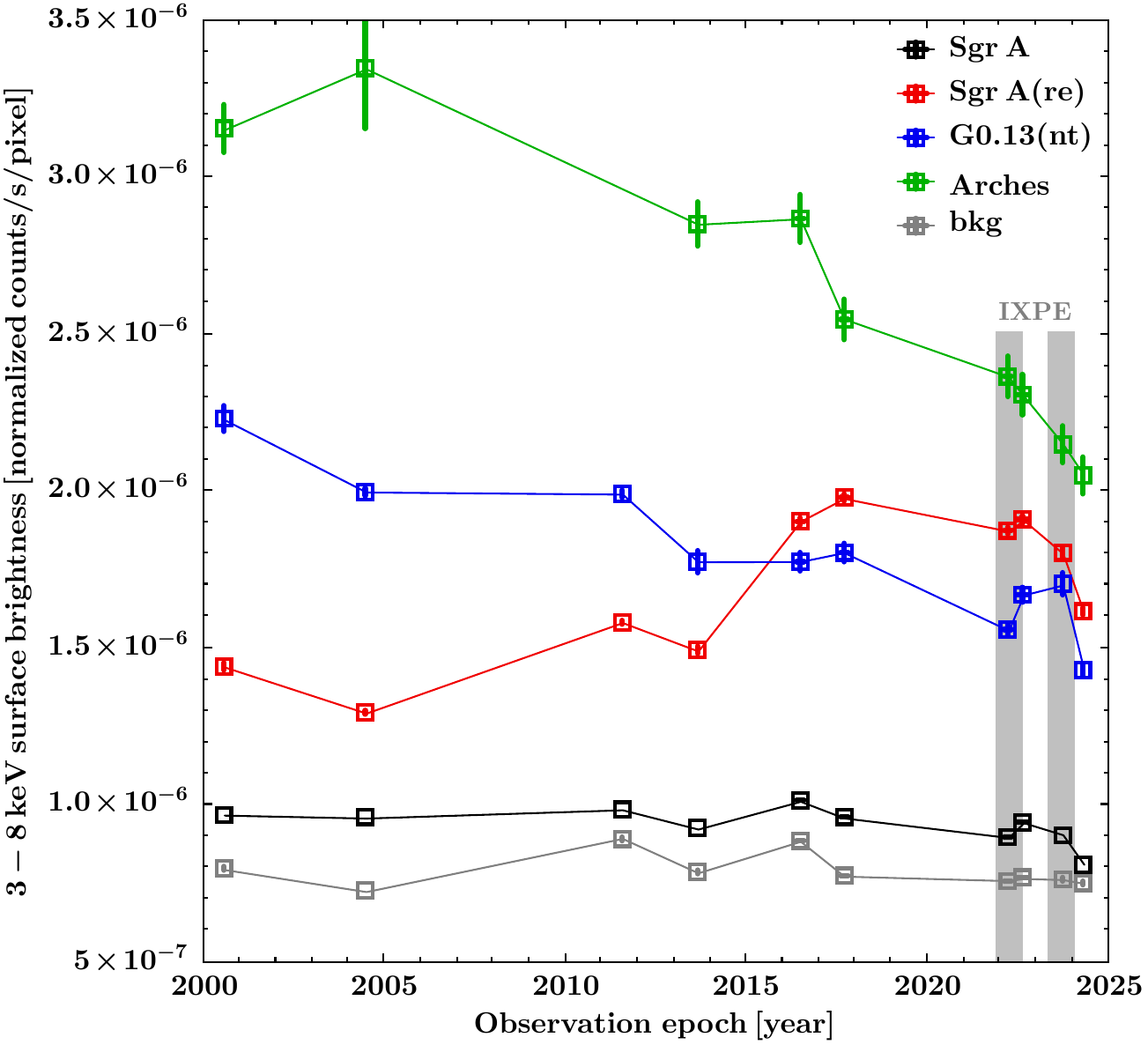}
\caption{Epoch-to-epoch variations in the  3--8 keV surface brightness {(normalized per $0.5\arcsec \times 0.5\arcsec$ pixels after subtraction of the particle background)} for the regions depicted \revision{in Fig.~\ref{fig:chandra_epochs}}. {Namely, this figure shows the light curves for a region encompassing the Arches cluster (green), {the central brightest region of the Sgr~A complex (red), the nonthermal (NT) nebula G0.13-0.11 (blue),} and the full Sgr~A complex excluding the last two subregions (black) in comparison to a nearby background region (gray). } The full time span of the observations is more than 20 years, and the time windows encompassing IXPE observations are shown as shaded boxes. For most of the data points, the statistical uncertainty of the flux measurement is smaller than the symbol size.
}
\label{fig:chandra_lc}
\end{figure}

\begin{figure}
\centering
\includegraphics[angle=0,trim=2cm 2cm 2cm 0cm,clip,width=1.\columnwidth]{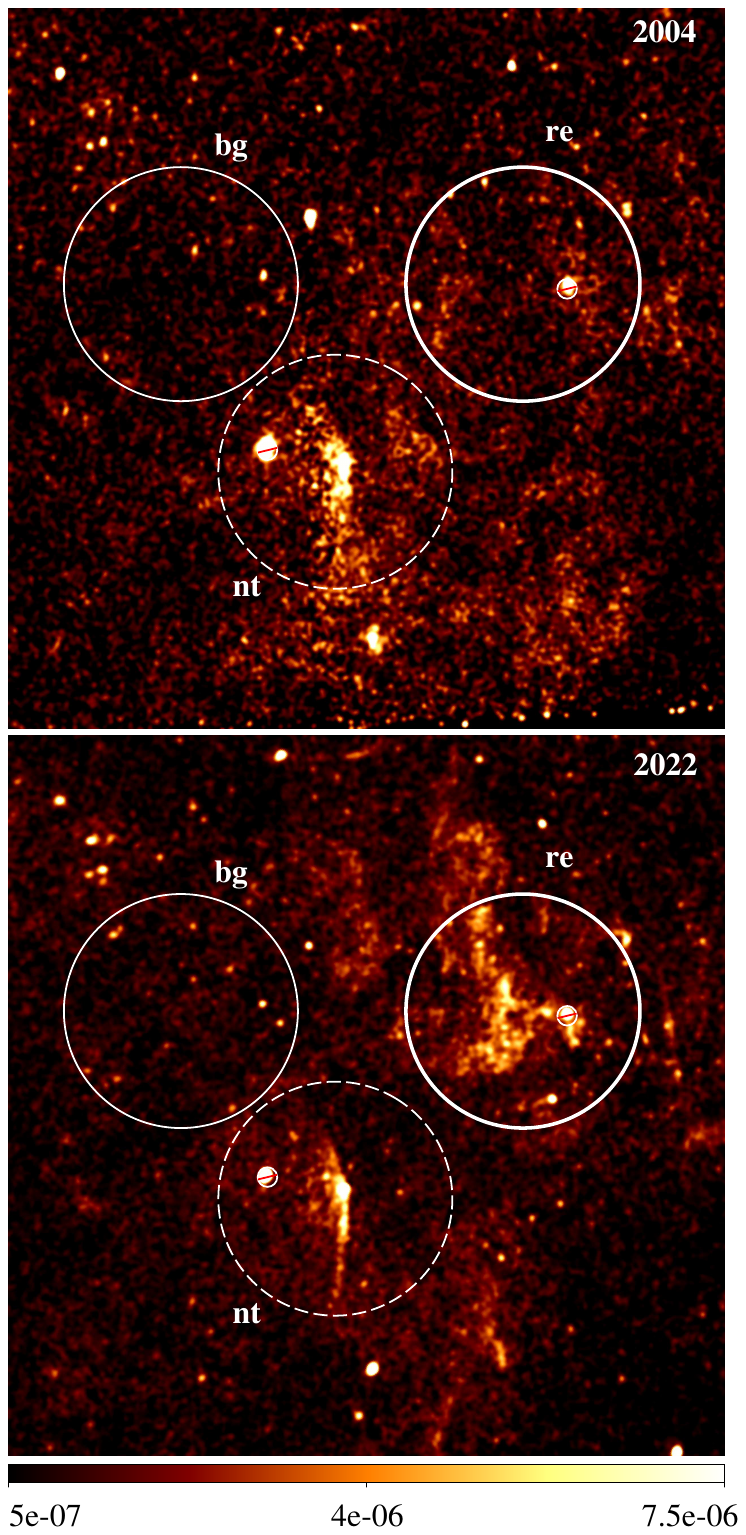}
\caption{Same as Fig.~\ref{fig:chandra_epochs} but showing a zoom-in of the interior part of the Sgr~A complex and focusing on the difference between two epochs, 2004 (top) and 2022 (bottom), to highlight the small-scale variations in the morphology of the diffuse emission. Big circular regions are 1$\arcmin$ in radius and are used for the differential spectral analysis (after masking the brightest point sources marked with small crossed circles). These regions correspond to the brightest reflection spot (marked "re"), a region dominated by nonthermal emission of the PWN G0.13-0.11 (marked "nt"), and a representative subregion of the Sgr~A complex (marked "bg").  
}
\label{fig:chandra_zoom}
\end{figure}

\begin{figure}
\centering
\includegraphics[angle=0,trim=0.5cm 5cm 1cm 3.7cm,clip,width=0.8\columnwidth]{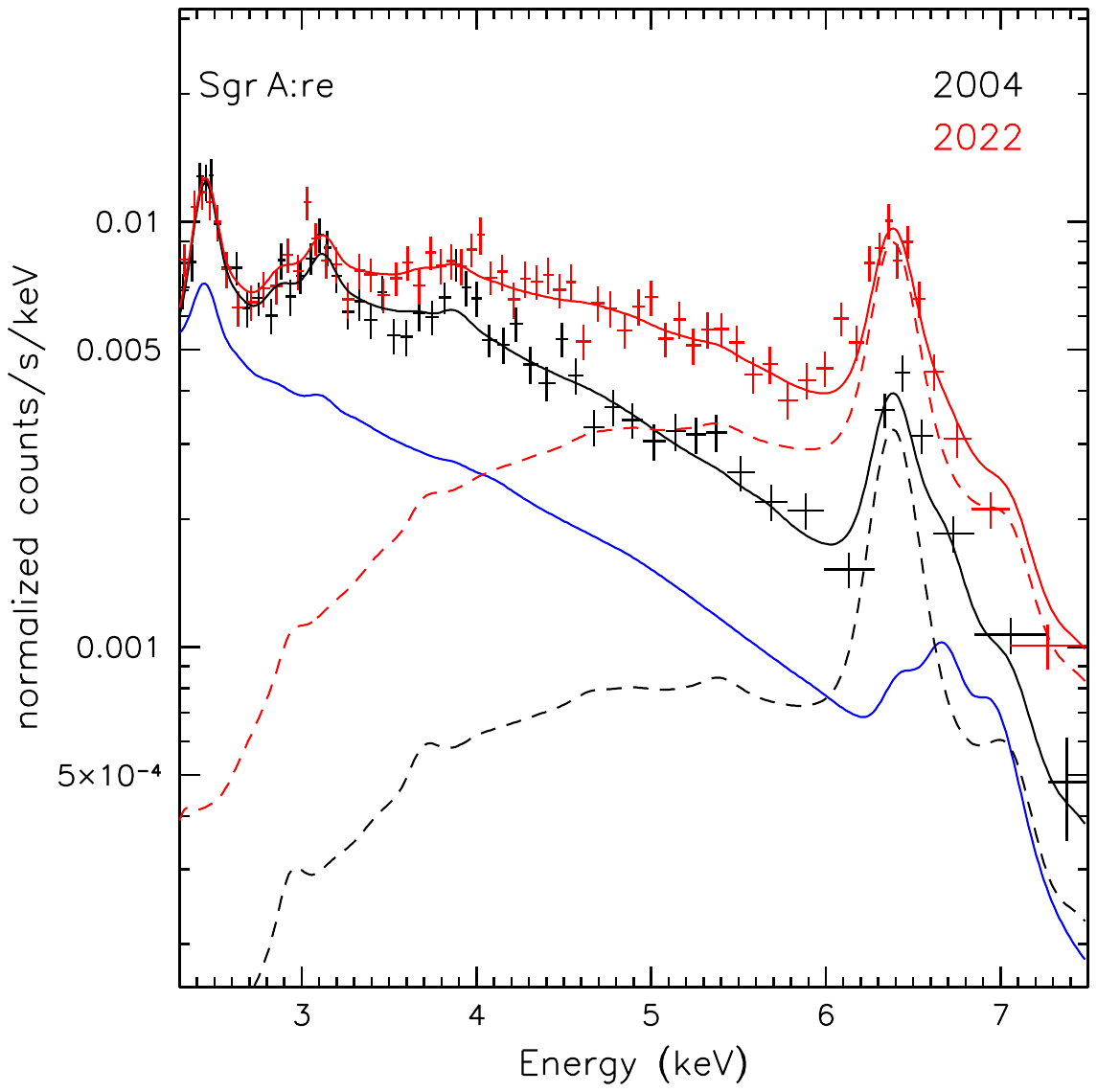}
\includegraphics[angle=0,trim=0.5cm 5cm 1cm 3.7cm,clip,width=0.8\columnwidth]{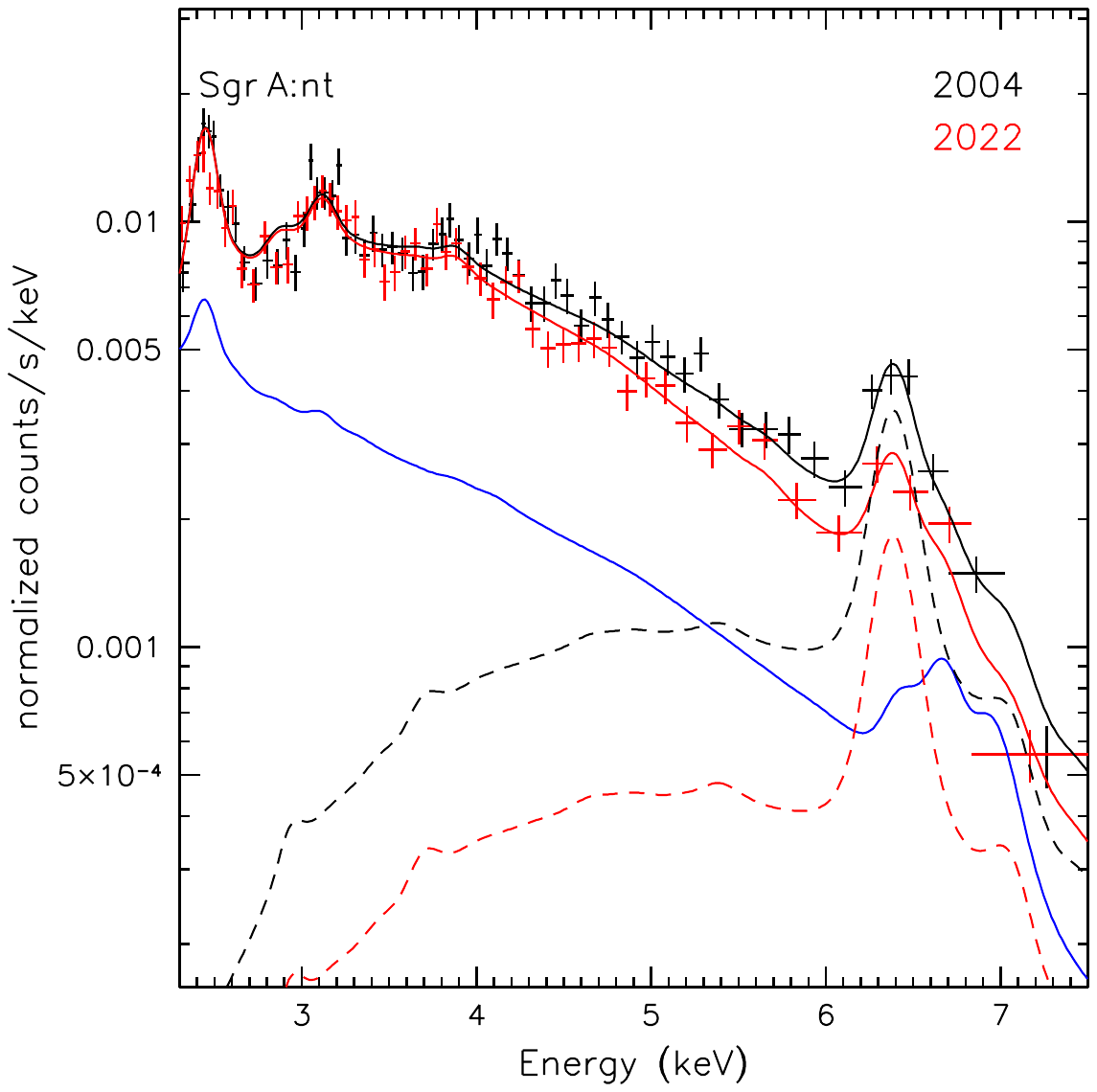}
\includegraphics[angle=0,trim=0.5cm 5cm 1cm 3.7cm,clip,width=0.8\columnwidth]{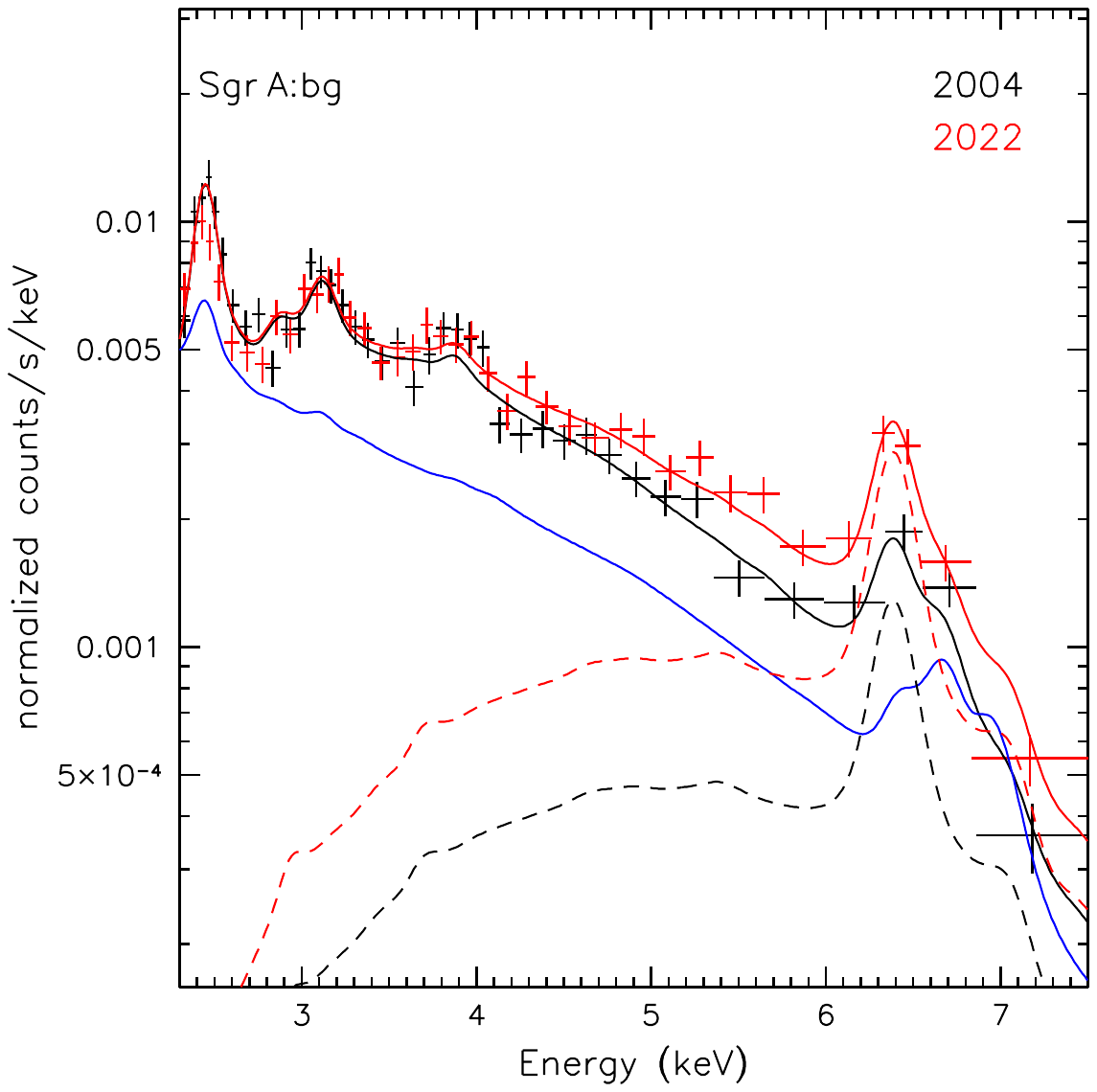}
\caption{Epoch-to-epoch variations in the spectra of several regions within the Sgr~A complex shown in Fig.~\ref{fig:chandra_zoom}. {Top panel: Brightest (in 2022) reflection region, "Sgr~A:re". Middle panel: Region around G0.13-0.11, "Sgr~A:nt". Bottom panel: Relatively faint region within the Sgr~A complex, "Sgr~A:bg".} The spectra taken from \textit{Chandra} observations in 2004 (black) and 2022 (red) are fit with the same model with all parameters tied except for normalizations of the reflected continuum and fluorescent lines. The composite reflection models for each observation are shown with dashed lines. The expected contribution of the spatially smooth Galactic bulge emission is shown in blue for comparison. {The data were binned only for visualization purposes.}
}
\label{fig:chandra_spectra}
\end{figure}

The X-ray images of the Sgr~A region in the  3--8 keV band obtained by \textit{Chandra} in different epochs starting from 2013 are shown in Fig.~\ref{fig:chandra_epochs}. Clear global and local changes in the morphology are visible over this 10-year-long period, even though the whole region stays relatively bright over its entire duration. This behavior is indeed expected, given that 10 years corresponds to the light travel distance of $\sim$3 pc (1.27 arcmin at the Galactic Center distance), which is bigger or comparable to the physical sizes of individual substructures ($\sim1$ pc), but smaller than the full complex itself ($\sim10$ pc). 

\subsection{Light curves}

Epoch-to-epoch light curves of the emission in the  3--8 keV band for several regions of this field are shown in Fig.~\ref{fig:chandra_lc} \revision{(exact definitions of the regions are listed in Table~\ref{t:regions})}. One can see that the average surface brightness of the large circular region encompassing the Sgr~A complex (black line) from 2000 to 2024 is relatively stable, even though the morphology of the emission within it was changing. Slight ($\sim10\%$ level) variations are present, some of which are caused by different positioning of this relatively large region in \textit{Chandra}'s FoV, resulting in different weighting of various parts of it due to spatially dependent vignetting and locations of the chip gaps, as well as the contribution of time-variable and transient point sources. This is illustrated by the presence of similar amplitude variation in a background region, which appears to contain a much smaller amount of reflected emission.

The other three regions are smaller in size and show substantial variations in surface brightness, with the brightest central part of the Sgr~A complex demonstrating a rapid rise from 2013 to 2015 and then a decay after 2023. Decay of the diffuse emission surrounding the Arches cluster is well known and is also (partially) attributed to the decay of the reflected emission \citep[e.g.,][]{2011A&A...525L...2C,2014ApJ...781..107K,2014MNRAS.443L.129C,2017MNRAS.468.2822K,2019MNRAS.484.1627K}. More surprising is a decrease in surface brightness from the region dominated by the nonthermal emission from G0.13-0.11 nebula. However, spectral analysis of the variable part of the diffuse emission shows that it is also the decay of the reflection emission that drives this apparent trend { (see Sect.~\ref{ss:spectra})}.

\subsection{Spectra}
\label{ss:spectra}

Spectroscopy of the time variable emission is a very powerful tool that allows us to get an idea about the spectral shape of the variable and non-variable components, which can then be used for single epoch spectral decomposition. In order to make this method most efficient, \revision{one can select two epochs of observations that correspond to the minimal and maximal 3$-$8 keV surface brightness in the regions of interest. We focus on three regions well within the Sgr~A complex shown in Fig.~\ref{fig:chandra_zoom} and consider the difference between observations taken in 2004 and 2022, optimizing for the brightest central region in Sgr~A (see Fig.~\ref{fig:chandra_lc}).}

We extract spectra in each of them and fit simultaneously by a model \citep[described and exploited for Sgr~A complex in ][]{2023Natur.619...41M}, allowing only the reflection component \citep[derived from \texttt{CREFL16} model\footnote{\url{https://wwwmpa.mpa-garching.mpg.de/~churazov/crefl}} by ][]{2017MNRAS.468..165C} to vary. The other components (`the hot ($kT\sim7$ keV) plasma' and `warm ($kT\sim 2$ keV) plasma') are essentially phenomenological descriptions of the temporally constant and spatially smooth emission of background and foreground sources { (see the model definitions in \citealt{2023Natur.619...41M}).} \revision{The spectral shapes of all components were fixed (according to the parameters found by analysis of sensitive epoch-resolved \textit{XMM-Newton} and \textit{Chandra} data of Sgr~A performed in \citealt{2023Natur.619...41M}) and only their normalizations were fitted using standard procedures of \textsc{XSPEC}\footnote{\url{https://heasarc.gsfc.nasa.gov/xanadu/xspec/}} software package \citep[][V12.14]{1996ASPC..101...17A}. Although the spectral shape of both the reflection signal and the contaminating emission might vary from region to region, the phenomenological model constructed for the full Sgr~A complex is flexible enough to describe these variations by changing the relative contributions of its components. In particular, in the case of the reflection signal, the \texttt{CREFL16} model is split into continuum (\texttt{CREFL$_{\rm cont}$}) and fluorescent line (\texttt{CREFL$_{\rm fluor}$}) components, and freedom in their relative normalization is capable of absorbing possible variations in the spectral shape of the reflected emission given limited energy range of the available spectra. The non-reflection components of the model do not play any role in the polarization analysis, given that their spectral shapes are sufficiently `orthogonal' to the spectral shape of the time-variable reflection components.  }

Very good quality of the fits ($\chi^2$=685 and 705 for 705 degrees of freedom for the "bg" and "re" reflection-dominated regions, and  $\chi^2$=740 with 705 degrees of freedom for the G0.13-0.11:"nt" region due to the presence of a nonthermal powerlaw component in this { case) proves} that the variable emission has the spectral shape that is indeed fully compatible with the reflection scenario (as shown in Fig.~\ref{fig:chandra_spectra}). In particular, the difference spectrum contains a hard X-ray continuum and a prominent fluorescent iron line that are treated separately during the fit to allow more flexibility in the spectral shape. The resulting non-reflection emission can be compared with the prediction of the spatial-decomposition model, which is capable of estimating the brightness of the bulge and ridge emission based on the projected stellar mass maps \citep[][]{2025A&A...698A.313A}. As a proxy for the spectral shape of this emission, we use spectra extracted from a deep observation of the bulge region \citep[][]{2009Natur.458.1142R} which is believed to contain negligible contribution of the diffuse reflected emission (even though featuring the 6.4 keV line due to the presence of such a line in spectra of certain types of accreting white dwarfs; { see, e.g., \citealt{2025A&A...698A.313A})}. One can see that the emission predicted by this model does not exceed the derived non-variable part of the spectrum for these regions, \revision{which we consider is a useful sanity check for the time variability-informed spectral decomposition.} 

We also observed that even for the region dominated by nonthermal power-law emission of the G0.13-0.11 nebula, the spectrum of the emission responsible for the drop in brightness since 2004 is fully consistent with the reflection model. Interestingly, the relative strength, or equivalent width, of the fluorescent line differs slightly from region to region. Although this might be an indication of a {different reflection geometry, it is also within expected variations due to inhomogeneities in iron abundance or the relative contribution of second scatterings} (which is expected to be larger for fainter regions and feature a stronger fluorescent line with a Compton shoulder profile).\footnote{Once the primary photons leave the cloud, 
the cloud fades, but the doubly and multiply scattered photons might still be detected with deep observations \citep[{ see, e.g.,}][]{1996AstL...22..648S,2002MNRAS.330..817C,2016A&A...589A..88M,2022MNRAS.509.1605K,2022MNRAS.509.6068K}.} In the case of abundance variations, the predicted polarization signal is unchanged, while the other two cases predict correlated change in the polarization degree \citep[{ see, e.g.,}][]{2002MNRAS.330..817C,2020MNRAS.498.4379K}. In these two cases, stronger line emission should correspond to a smaller polarization degree of the continuum, either as a result of contamination by less polarized second scattering emission or due to the phase function of the Compton scattering {(for the spectral band of interest here, Thomson limit of the Compton scattering provides a fully adequate description)} as a function of the cosine of the scattering angle. Another cause of spatially variable equivalent width of the fluorescent line might be intrinsic polarization of the primary emission, which in principle allows one to produce a high polarization degree in the continuum and a relatively strong emission line, but only at the cost of a relatively low continuum emission overall \citep[][]{2020MNRAS.498.4379K}. In this case, the direction of the polarization vector might be very different from the unpolarized primary source prediction. 

Imaging, timing, and spectral analysis of the \textit{Chandra} data allowed us to quantify reflected continuum properties and predict the polarization signal we could expect as a function of the reflection geometry parameters. We are now in a position to combine this with the IXPE data to derive geometrical constraints based on the spatially resolved spectropolarimetric measurements.

\section{Spatially resolved X-ray spectropolarimetry} 
\label{s:polarimetry}

The angular and spectral resolution of IXPE, in addition to the limited time coverage of the Galactic Center region, is not sufficient to disentangle different emission components contributing to the observed signal in an independent manner. We have to rely on spectral and spatial information derived from quasi-simultaneous \textit{Chandra} observations combined with the historical  \textit{Chandra} and \textit{XMM-Newton} data, setting the baseline for the non-variable diffuse emission components. In this regard, the first step needed to be done is checking the consistency of the total (i.e., not component-separated) imaging and spectral properties in IXPE Stokes $I$ and \textit{Chandra} data. 

\subsection{IXPE Stokes $I$ images and spectra}
\label{s:ixpespectra}
\begin{figure*}
\centering
\includegraphics[angle=0,trim=1.8cm 7cm 2.5cm 5cm,clip,width=0.66\columnwidth]{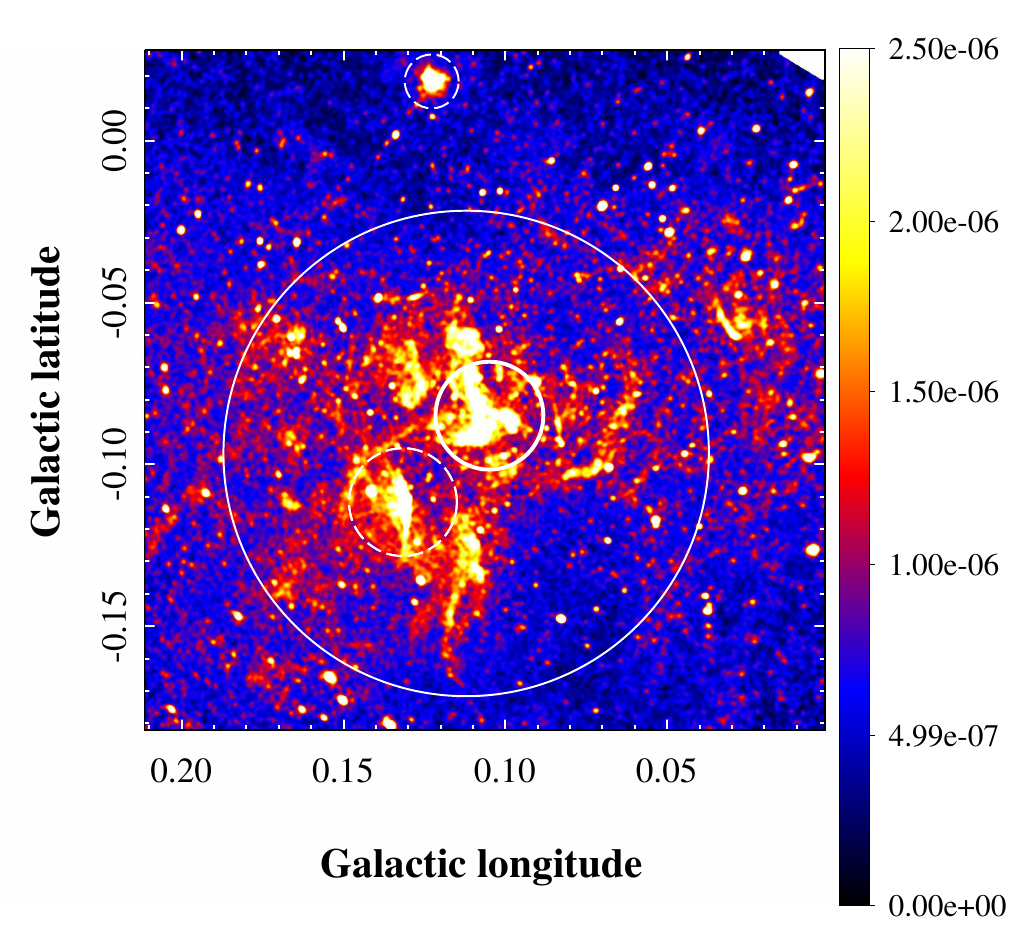}
\includegraphics[angle=0,trim=1.8cm 7cm 2.5cm 5cm,clip,width=0.66\columnwidth]{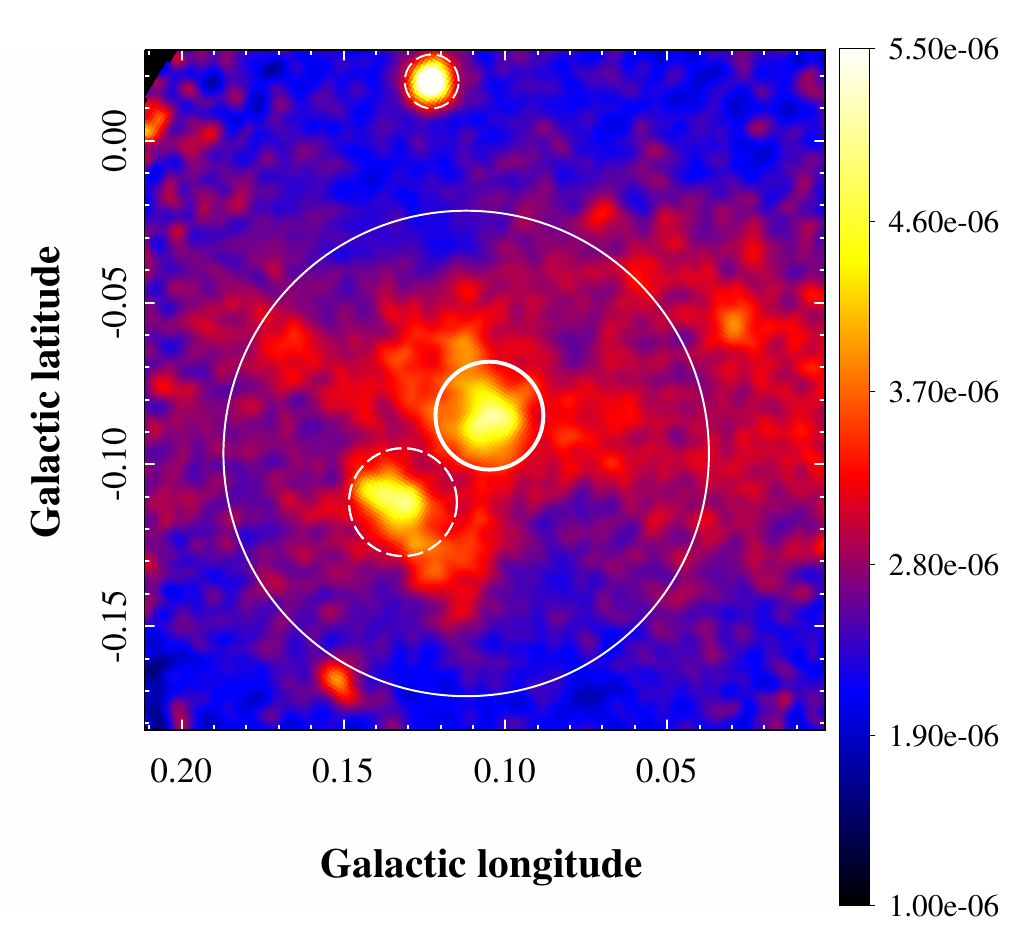}
\includegraphics[angle=0,trim=1.8cm 7cm 2.5cm 5cm,clip,width=0.66\columnwidth]{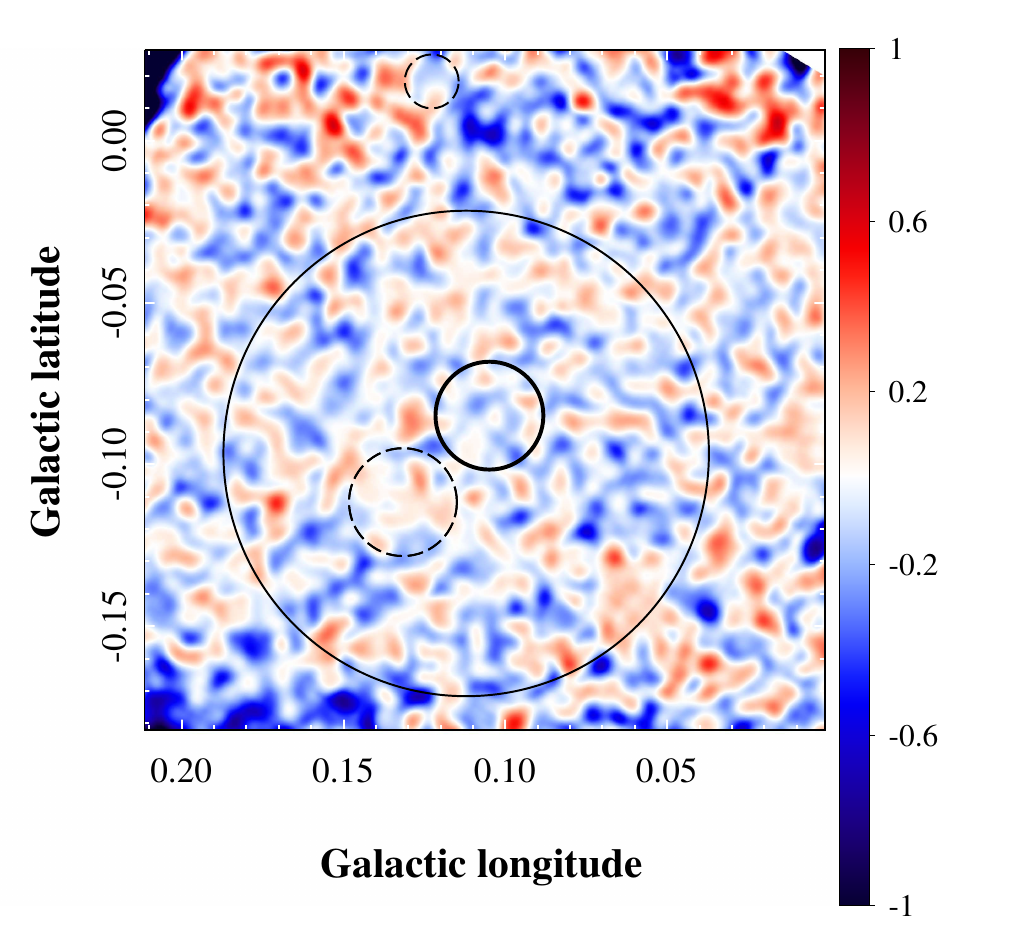}
\caption{Comparison of the \textit{Chandra} (2022, left) and IXPE (combined 2022 and 2023 datasets, middle)  3--8 keV images (${\rm ct~s^{-1}~pixel^{-1}}$, linear scale) and relative difference between the two \textit{Chandra} observations in 2022 and 2023 (right panel).  Very good epoch-to-epoch consistency is indicated by the absence of significant structures in the difference image, while the similarity of the \textit{Chandra} and IXPE images shows the good consistency of the imaging by the two telescopes.
}
\label{fig:chandraixpeimages}
\end{figure*}
\begin{figure}
\centering
\includegraphics[angle=0,trim=0.7cm 5cm 0.5cm 3.5cm,clip,width=0.85\columnwidth]{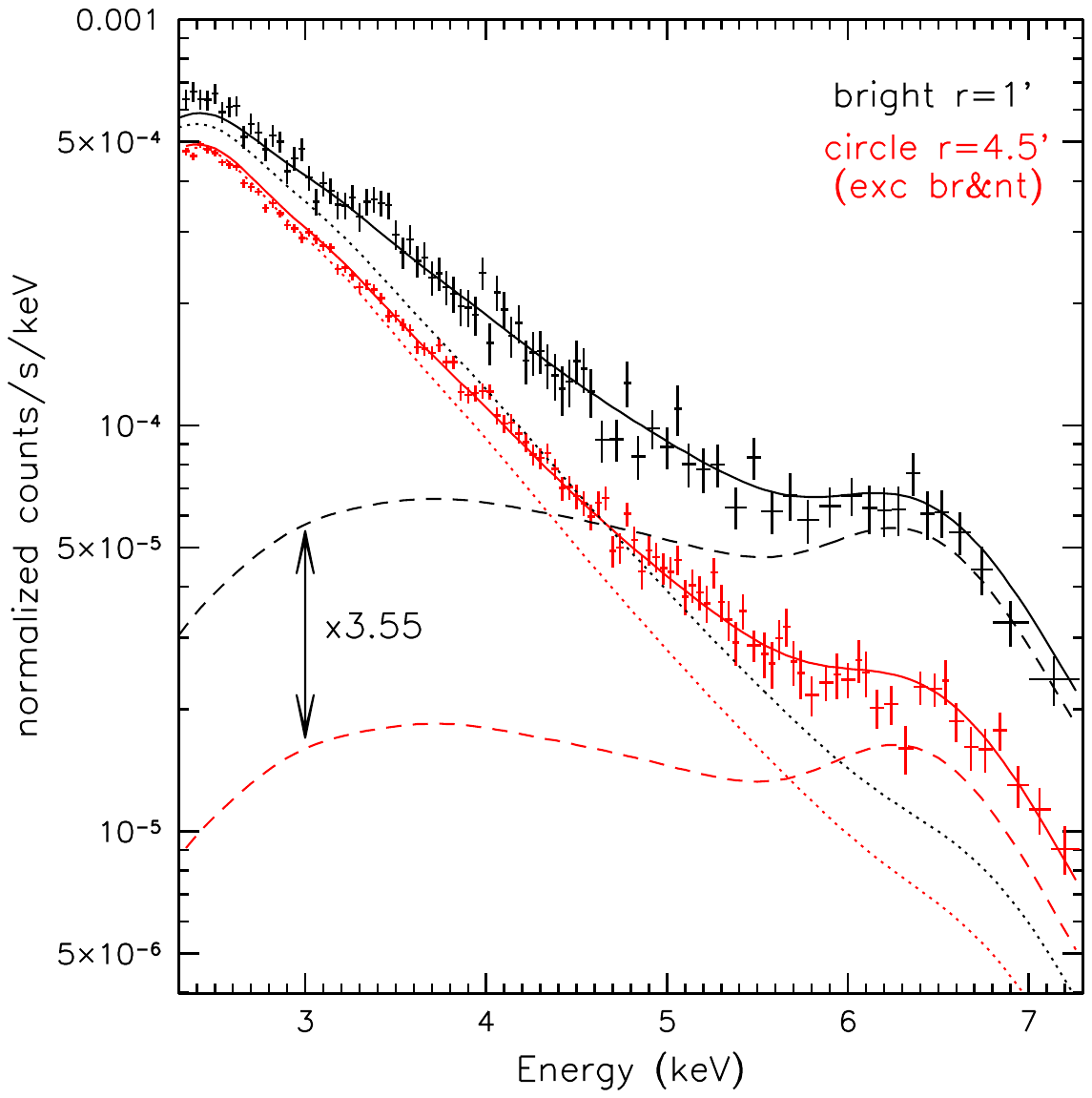}
\caption{Spectra from IXPE for the two regions used for the polarimetric measurements of the reflected emission. The solid lines show the total spectral models inferred from the \textit{Chandra} data, while dashed and dotted lines show contributions of the reflection and non-reflection emission for each 
region. The surface brightness of the reflected continuum in the brightest region is 3.55 times higher than in the large circle region. {The data were slightly binned (in the high-energy part) for visualization purposes only.}
}
\label{fig:ixpe_spectra}
\end{figure}

Figure~\ref{fig:chandraixpeimages} shows a comparison of \textit{Chandra} and IXPE  3--8 keV images for the region of interest. The \textit{Chandra} image corresponds to the middle 2022 epoch, while the IXPE image combines both 2022 and 2023 observation epochs. Only small morphological variations between 2022 and 2023 observations are expected from a comparison of the \textit{Chandra} images taken in these epochs. The consistency of \textit{Chandra} and IXPE images is very good { (see Fig.~\ref{fig:chandraixpeimages})}, both in the large-scale diffuse emission and for the three brightest individual regions -- the Arches cluster, the G0.13-0.11 PWN, and the brightest reflection spot in the Sgr~A complex. Some differences in morphology or relative brightness of certain structures are expected, given that the energy dependencies of the effective area of the two telescopes are different, resulting in slightly different weighting of various emission components in the combined  3--8 keV image. 
{In what follows, we consider the combined IXPE dataset and demonstrate consistency of differences between the epochs with the expected statistical deviation level in Appendix~\ref{s:systematic}.
}

The consistency of the IXPE Stokes $I$ and \textit{Chandra} data can be further validated by comparing \revision{solid angle-normalized} spectra for several regions of interest. Figure~\ref{fig:ixpe_spectra} shows the IXPE Stokes $I$ spectra \revision{(corresponding to the surface brightness per arcmin$^2$)} for two mutually exclusive regions: a large (r=4.5$\arcmin$) circle region encompassing Sgr~A (also excluding the region around the G0.13-0.11 nebula) and the brightest reflection spot in its center. Both \revision{area-normalized} spectra are perfectly fit with the models derived from the quasi-simultaneous \textit{Chandra} data with only a slight normalization correction (factor 0.85) needed, which is well within the possible margin of the effect introduced by proper response function calculations for extended sources with complicated morphological structure. This factor is propagated further in a self-consistent manner, so the polarization degree calculations are not affected by this.

Figure~\ref{fig:ixpe_spectra} also shows, separately, contributions of the reflected (dashed lines) and non-reflected (dotted lines) emission. It is the continuum part of the reflection continuum that is expected to be polarized, and these reflection continuum models are used in the spectropolarimetric analysis later on. One can see that the reflection continuum component of the brightest region is 3.55 times brighter for the central region compared to the average over the large circle. Since the fluorescent line emission of the reflection component should not be polarized, we focus the polarimetric analysis on the 3--6 keV band, which maximizes signal-to-noise ratio for the reflection continuum.
 {We summarize parameters of the reflection models for these regions and their relative contributions to the 3--6 keV band in Table~\ref{t:spec}. }

\begin{table*}
\caption{Parameters of the reflection model \texttt{TBabs(CREFL16$_{\rm cont}$+CREFL16$_{\rm fluor}$)}.}
\begin{center}
\begin{tabular}{lcccccccccc}
\hline
{ Region} & $N_{\rm H}$ & \multirow{2}{*}{$\Gamma$} & \multirow{2}{*}{$\tau_{\rm T}$} & \multirow{2}{*}{$\mu$} & \multirow{2}{*}{\revision{$A$}} & Norm$_{\rm cont}$ & Norm$_{\rm fluor}$ & \multirow{2}{*}{Frac$_{3-6,\rm cont}$}& \\ 
       & $10^{22}\,\rm cm^{-2}$ & & & & & $10^{-2}$~per arcmin$^2$& $10^{-2}$ per arcmin$^2$&  \\
\hline
 Sgr~A, $r=4.5\arcmin$ (excl. br\&nt) &  \multirow{2}{*}{7.0} &  \multirow{2}{*}{$1.8$} & \multirow{2}{*}{ $13.0$} & \multirow{2}{*}{$-$0.3} & \multirow{2}{*}{1.0} & $1.7\pm0.1$ & $1.2\pm0.1$ & 0.24\\
Sgr~A, bright $r=1\arcmin$ &   &   &  &  &  & $5.9\pm0.25$ & $3.6\pm0.25$ &0.46 \\
\hline
\hline
\end{tabular}
\end{center}
\tablefoot{\revision{The spectral shape was derived from spectral analysis {of the time-variable diffuse emission} in \textit{Chandra} data and only  normalizations were adjusted to best fit the full spectra (together with other components).} Normalization was adjusted by a constant factor 0.85 to match IXPE Stokes $I$ spectra for the same regions. The columns show absorbing column density ($N_{\rm H}$), incident powerlaw index $\Gamma$, optical depth of the cloud $\tau_{\rm T}$, cosine of the scattering angle $\mu$, abundance of heavy elements with respect to the Solar values \revision{$A$} (see \citealt{2017MNRAS.468..165C} for a detailed description of the \texttt{CREFL16} model), \revision{and they were fixed at values found in combined epoch-resolved \textit{XMM-Newton} and \textit{Chandra} spectroscopy performed in \citealt{2023Natur.619...41M}}. The normalizations (defined per arcmin$^2$) of the continuum (Norm$_{\rm cont}$) and of the fluorescent lines (Norm$_{\rm fluor}$) components of this model were mutually untied and independently fitted for the regions of interest here (the uncertainties correspond to 1$\sigma$ confidence intervals). {Intrinsic model's normalization} equal to $10^{-2}$ corresponds to unabsorbed  3--8 keV flux of $9.1\times10^{-14}$ ${\rm erg~s^{-1}~cm^{-2}}$ and $4.2\times10^{-14}$ ${\rm erg~s^{-1}~cm^{-2}}$ for the continuum and fluorescent components respectively. The last column, Frac$_{3-6,\rm cont}$, gives the relative contribution of the reflected continuum to the total 3--6 keV flux from these regions.}
\label{t:spec}
\end{table*}

\subsection{IXPE Stokes $Q$ and $U$ images and spectra}
\begin{figure}
\centering
\includegraphics[angle=0,trim=1.8cm 7cm 2.5cm 5cm,clip,width=0.96\columnwidth]{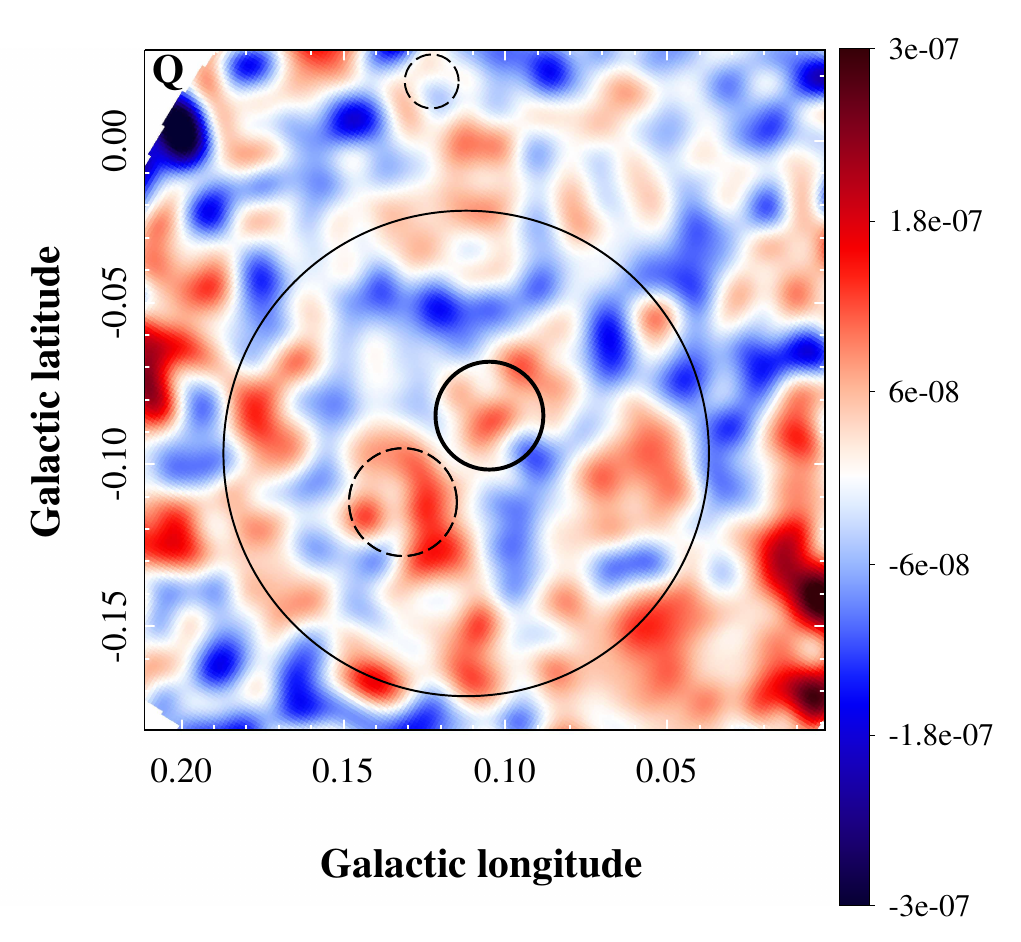}
\includegraphics[angle=0,trim=1.8cm 7cm 2.5cm 5cm,clip,width=0.96\columnwidth]{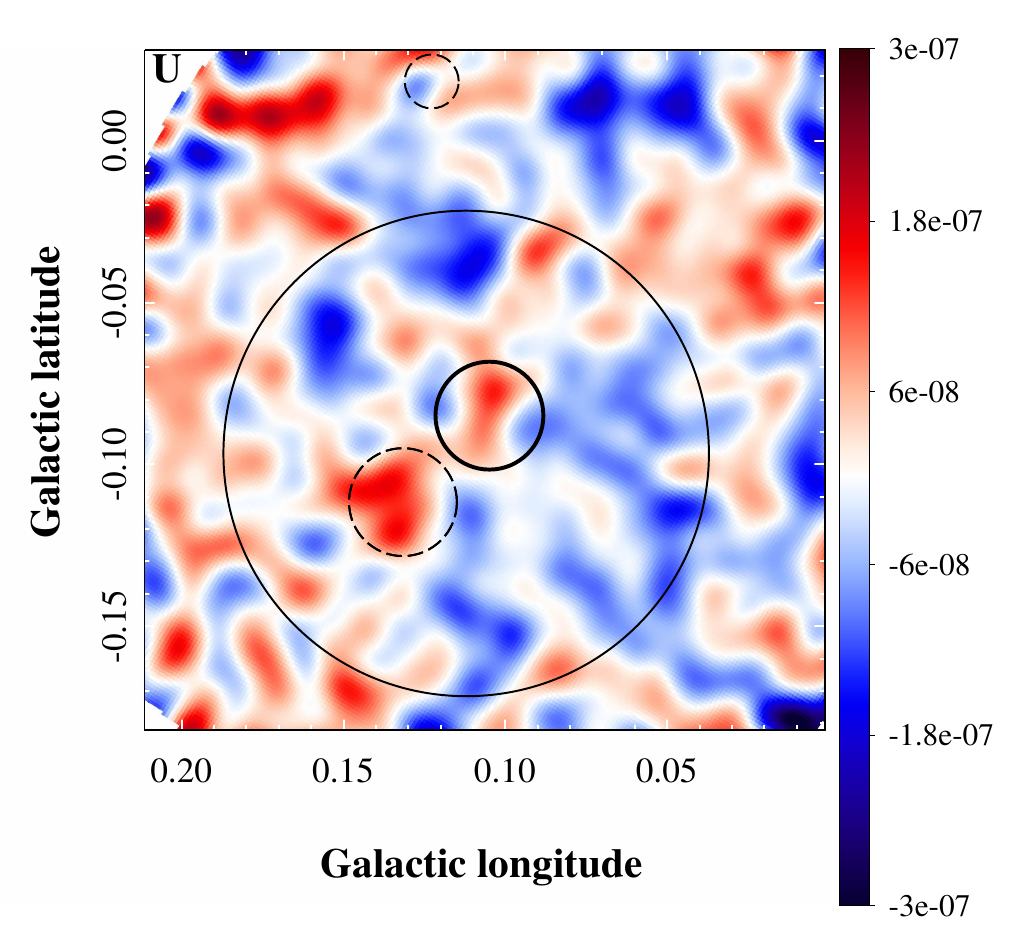}
\caption{Maps of the Stokes parameters $Q$ (top) and $U$ (bottom) in the 3--6 keV band (${\rm ct~s^{-1}~pixel^{-1}}$, linear scale) smoothed with a Gaussian kernel with $\sigma=30''$. {The same regions as before are highlighted (and marked in the top panel).}
}
\label{fig:stokesimages}
\begin{picture}(0,0)
\put(-54,488){\rotatebox{0}{\small \textcolor{black}{Arches}}}
\put(-17,465){\rotatebox{0}{\small \textcolor{black}{r=4.5'}}}
\put(-12,428){\rotatebox{0}{\small \textcolor{black}{r=1'}}}
\put(-30,405){\rotatebox{0}{\small \textcolor{black}{nt}}}
\end{picture}
\end{figure}
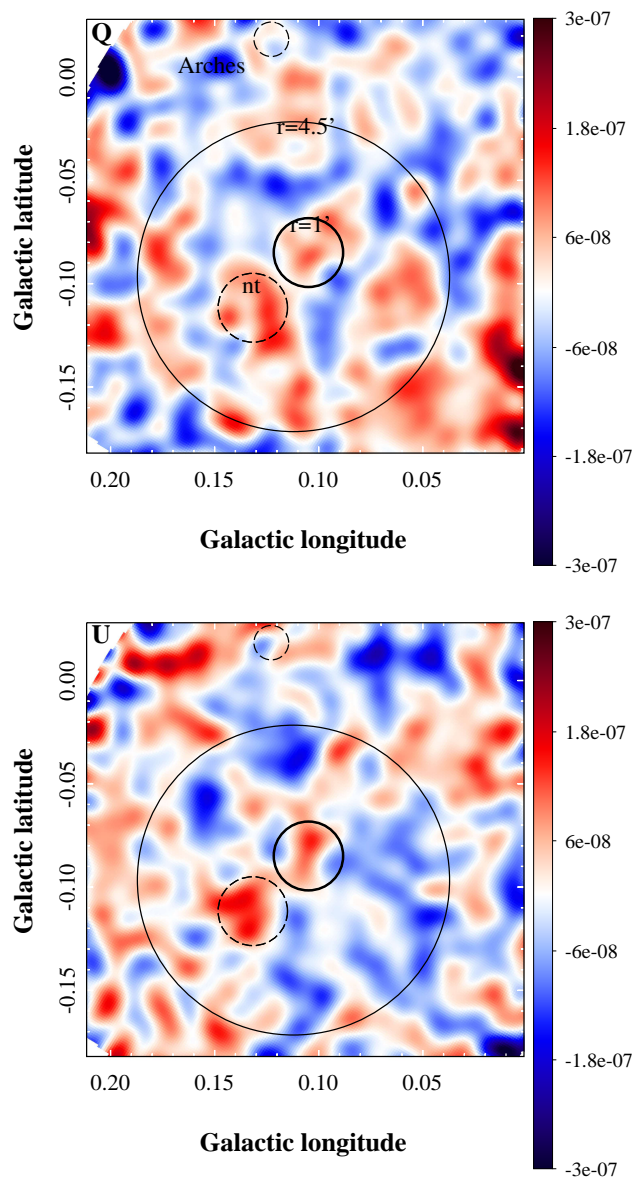
\begin{figure}
\centering
\includegraphics[angle=0,trim=0.7cm 5cm 0.5cm 3.5cm,clip,width=0.85\columnwidth]{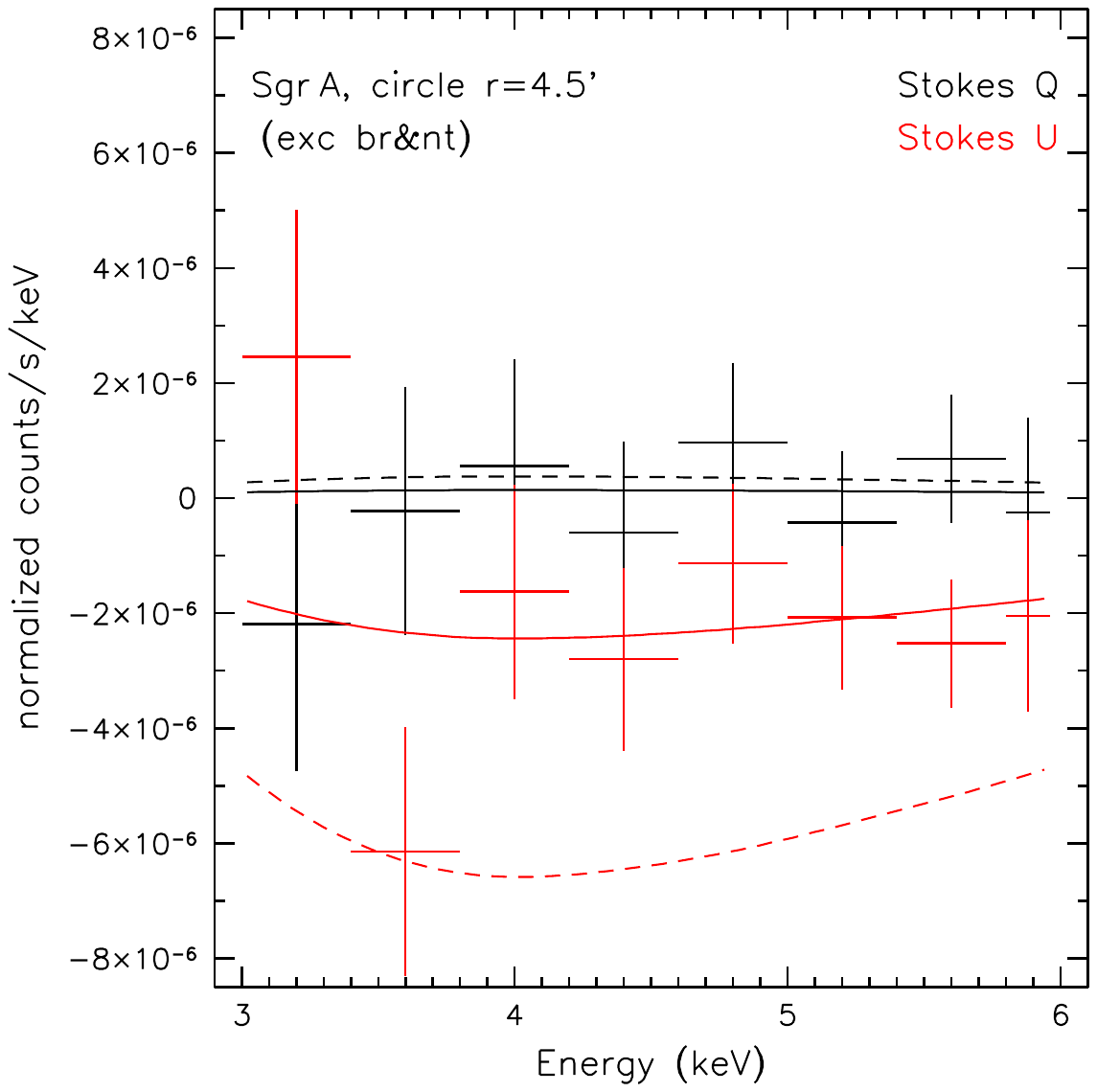}
\includegraphics[angle=0,trim=0.7cm 5cm 0.5cm 3.5cm,clip,width=0.85\columnwidth]{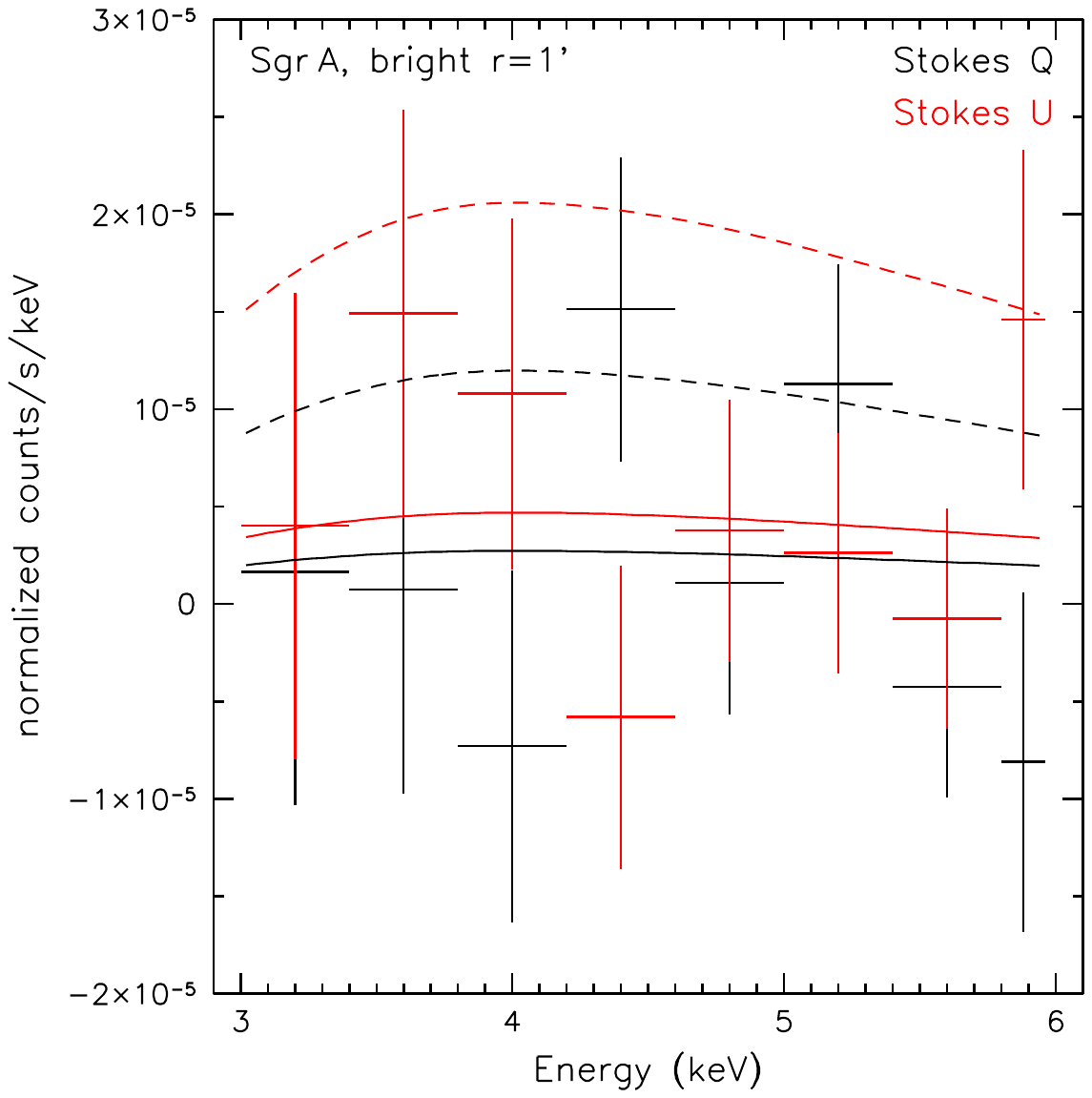}
\caption{IXPE spectra of the Stokes $Q$ and $U$ parameters for the two (mutually exclusive) regions used for the polarimetric measurements of the reflected emission: the $r=$4.5$\arcmin$ circle (also excluding regions around G0.13-0.11) and the brightest reflection region. The dashed lines show the expected $Q$ and $U$ spectra for the model with the same PA and Stokes $I$ and PD=100\%. {The data were uniformly binned only for visualization purposes.}
}
\label{fig:stokesspectra}
\end{figure}

\begin{figure}
\centering
\includegraphics[angle=0,trim=1.7cm 12.5cm 1.7cm 3.8cm,clip,width=0.85\columnwidth]{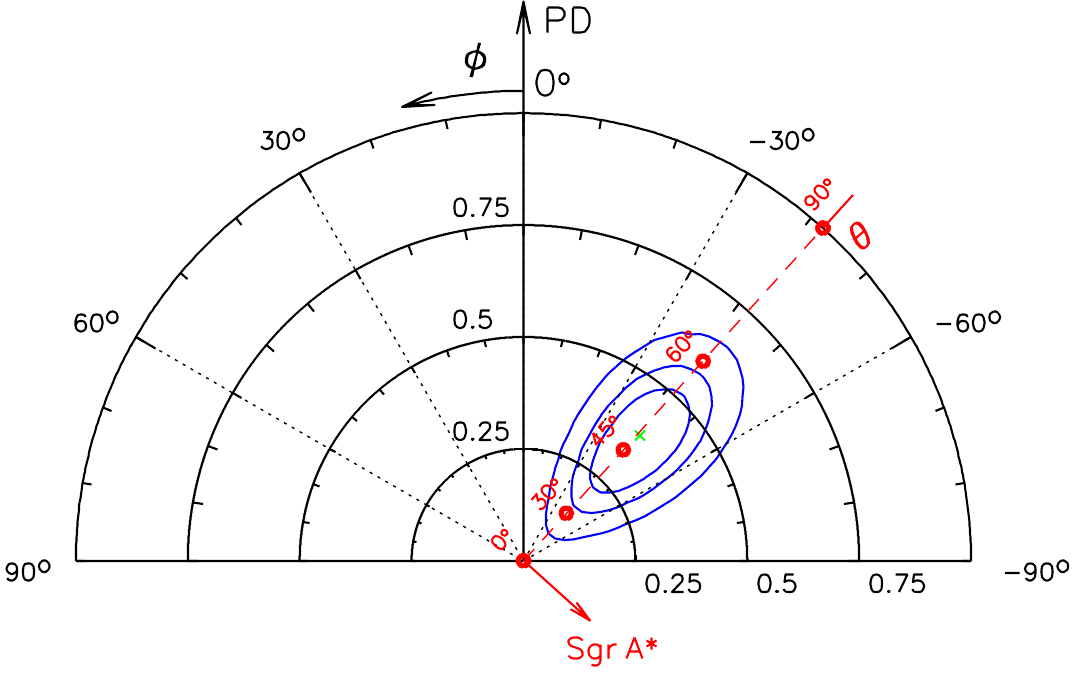}
\includegraphics[angle=0,trim=1.7cm 12.5cm 1.7cm 3.8cm,clip,width=0.85\columnwidth]{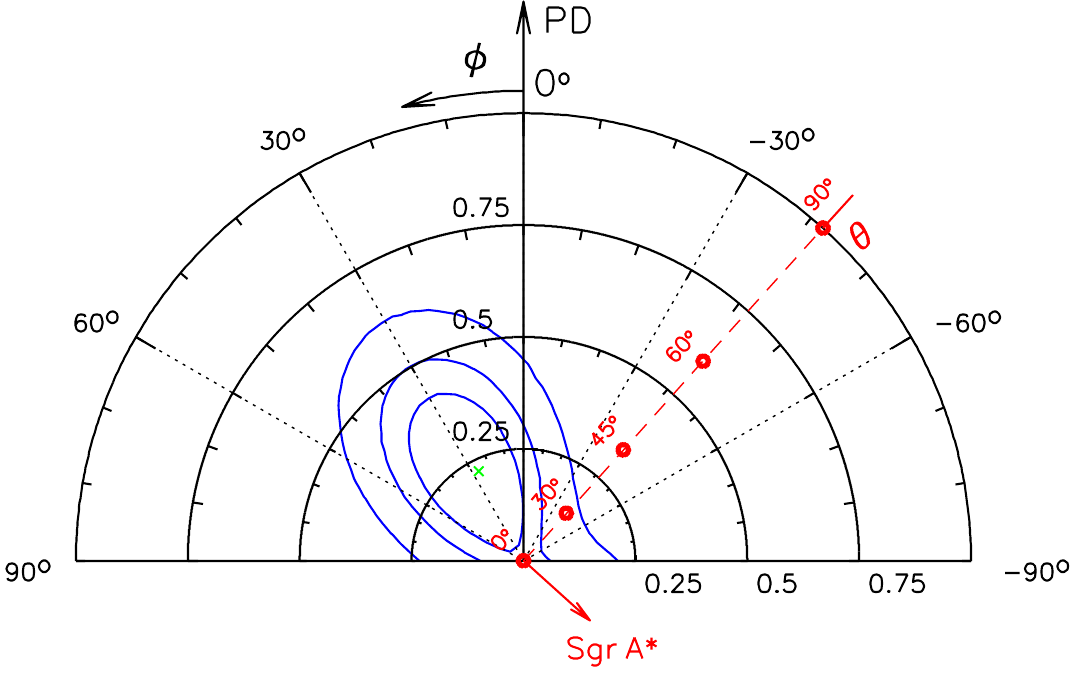}
\caption{Fitted parameters of the PA, $\phi$, and PD for the $r=$4.5$\arcmin$ circle (top) and the brightest $r=$1$\arcmin$ (bottom) regions. The contours show $\delta\chi^2$ levels of 2.3, 4.61, and 9.21 with respect to the best-fit parameters (marked with the green cross), corresponding to 68\% (1$\sigma$), 90\%, and 99\% confidence levels for the two fitted parameters. The direction toward Sgr~A* is also indicated as well as perpendicular to it \revision{(dashed line)}. The PDs corresponding to scattering angles $\theta=30,45,60$, and 90 degrees are also indicated.
}
\label{fig:stokesfits}
\end{figure}

Given the consistency of the morphological and spectral properties of the total intensity (Stokes $I$) of the diffuse emission measured by IXPE, we are in a position to proceed with mapping and spectral analysis of the polarization signal itself, which is most conveniently expressed in terms of the Stokes $Q$ and $U$ parameters. These are inferred for each individual photon (event), detected by IXPE, and the expectation value for an unpolarized source (point-like or diffuse) is zero (in the absence of systematic errors).

Linearity of the Stokes parameters allows one to produce maps in (optimal) energy bands and smooth or bin them in order to suppress statistical noise.  In Fig.~\ref{fig:stokesimages}, we show Stokes $Q$ and $U$ images obtained from the full IXPE dataset in the 3--6 keV band, where the signal-to-noise ratio for the reflected continuum is expected to be the highest. One can see dominance of the positive $Q$ and $U$ for the G0.13-0.11 region, close to zero $Q$ and negative $U$ for the large Sgr~A region, and marginally positive $Q$ and $U$ for the brightest reflection region. Also, the level of fluctuations due to background noise (and its increasing trend toward the boundaries) is visible. The latter effect is well known and described, arising from imperfections of the photoelectron tracks reconstruction near the boundaries of the FoV \citep[e.g., ][]{2023AJ....165..143D}. { Our choice of the extraction regions ensures that the possible impact of this effect is minimal, given that all {regions} are a) always located close to the centers of the detectors; b) are large enough to fully encompass diffuse structures which are bright in reflected emission (even taking into account the effective PSF of IXPE).} We try to quantify the possible amplitude of other systematic effects in Appendix~\ref{s:systematic}.

The described picture is consistent with the previous findings for the large Sgr~A region and the nonthermal emission from G0.13-0.11, but the result obtained for the brightest reflection region is very surprising. Indeed, the Stokes $Q$ and $U$ pattern for it appears more similar to the G0.13-0.11 region than to the Sgr~A complex, meaning the direction of polarization is almost perpendicular to the expected one.  

We proceed with the spectropolarimetric analysis in order to quantify and explore this in more detail. Following the approach used in \cite{2023Natur.619...41M} and \cite{2024A&A...686A..14C}, we construct energy spectra of the Stokes values $Q$ and $U$ extracted from the regions of interest and supplemented with corresponding response matrices. In addition to the effective area of the mirror system and efficiency of the detector system, the response matrices take into account the energy-dependent modulation factor, i.e., efficiency of the response to a linearly polarized signal.

Since we expect only the continuum part of the reflected emission component to be polarized, we build a model { \citep[][]{2023Natur.619...41M,2024A&A...686A..14C}} that allows one to simultaneously fit Stokes $Q$ and $U$ data, varying polarization degree (PD) and polarization direction (in terms of the PA $\phi$) for a fixed spectral shape and normalization of the reflected continuum for each region { (found consistently in \textit{Chandra} and IXPE Stokes $I$ spectral analysis described in Sect.~\ref{ss:spectra} and \ref{s:ixpespectra})}.  The fitting is done { via a standard \texttt{steppar} procedure in} in \texttt{XSPEC} which involves calculating statistics on an extensive grid of PD and $\phi$ values ranging from 0 to 100\% { (defined with respect to the spectral model of the reflection continuum)} and from $-90$ to 90 degrees (in IAU convention), respectively. { Namely, the model is composed of two additive components with tied spectral shape (given by the spectrum of the reflected continuum) and normalization (set by PD) multiplied by $\cos 2\phi $ (for Stokes $Q$) and $\sin 2\phi$ (for Stokes $U$) user-defined functions.} The results of this procedure are shown in Fig.~\ref{fig:stokesspectra}. {As already hinted by the maps of the Stokes parameters Q\&U, the signal from the large Sgr~A region is characterized by negative $U$ and close to zero $Q$ values, while for the brightest reflection spot, both $Q$ and $U$ are marginally positive with large statistical noise.}

\subsection{Inferred polarization parameters}
\label{ss:inferredpolar}

The inferred confidence regions for the large circle (excluding the G0.13-0.11 and the brightest central region) and the brightest central region are shown in Fig.~\ref{fig:stokesfits} along with the expectations for the reflection scenario with Sgr~A* being the primary source (which sets PA, i.e., $\phi$ in this plot) and a range of the scattering angles (which sets the PD). The contours mark $\delta\chi^2$ levels of 2.3, 4.61, 9.21 with respect to the best-fit parameters corresponding to 68\%, 90\%, and 99\% confidence levels for two fitted parameters.

The best fit values for the large region are ${\rm PD}=(37\pm9)\%$ and $\phi=-43\pm7$ deg (quoted uncertainties correspond to $1\sigma$ confidence interval), fully consistent with the first measurement reported in \cite{2023Natur.619...41M}, who found ${\rm PD}=(31\pm11)\%$ and $\phi=-48\pm11$ deg. The detection of nonzero polarization is at the level of 3.9$\sigma$ significance based on the combined dataset. A slight increase in the PD is associated with the exclusion of the brightest reflection region, which appears to have almost orthogonal polarization direction (if any; see below),
{even though the polarization parameters inferred for the large region with and without exclusion of the brightest reflection spot are consistent within the uncertainties \revision{we list polarization parameters measured for both cases in two different epochs in Appendix~\ref{s:systematic} for direct comparison}.}
Since this exclusion is made based on intensity only, without any prior knowledge of its polarization properties, no intentional bias is expected to be introduced by this procedure. 

Our results for the large region signify the direction to the primary source, fully compatible with the Sgr~A* scenario, and the age of the flare is now constrained to $190\pm25$ yr in this case (corresponding to the cosine of the scattering angle $\mu=\sqrt{\frac{1-PD}{1+PD}}=-0.68\pm0.07$ (the solution branch corresponding to positive $\mu$ would correspond to the age of $\approx35\pm5$ yr, which is unlikely given historically available observations of the Galactic Center region in X-rays).

As already hinted by the Stokes maps, the polarization signal extracted from the brightest reflection region is rather different. The best fit value for the PA is $\phi=30\pm16$ deg and polarization degree ${\rm PD}=(23\pm13)\%$. That is, the significance of the detection of nonzero polarization falls just short of the two sigma level. On the other hand, the same polarization parameters as inferred for the large Sgr~A region can be excluded with a confidence level of more than 99\%, given that the best-fit polarization directions are almost perpendicular to each other. The 99\% upper limit on the PD for the PA fixed at the direction toward Sgr~A* is ${\rm PD}_{99}=19\%$ (1$\sigma$ upper limit is ${\rm PD}_{1\sigma}=4\%$), meaning that a situation with Sgr~A* as the primary source and the cloud as significantly more distant (larger $\mu^2$) is still possible.

Given that the brightest reflection region appears to be encompassed by more extended diffuse emission with different polarization, one may assume that there is a contribution of the large-scale emission to the extracted signal from the brightest reflection region. In such a case, one would need to subtract (or model) this contribution when fitting the polarization spectra. Using the signal from the large region as an estimate for the background for the brightest reflection region, the fitting procedure results in ${\rm PD}=(44\pm18)\%$ (i.e., 2.4$\sigma$) and $\phi=35\pm12$ deg. Naturally, this increase in significance is driven partially by the fact that polarization in the background region is almost perpendicular to the marginally detected polarization in the brightest reflection region. The 99\% upper limit on the PD for the PA fixed at the direction toward Sgr~A* is ${\rm PD}_{99}=15\%$ (1$\sigma$ upper limit is ${\rm PD}_{1\sigma}=3\%$) in this case.

 {Finally, one might expect the impact of systematic effects for the brightest region in the center of the FoV. Analysis of the individual datasets (two observational epochs and 3 DUs) does not reveal exceptional outliers, although a significant detection is achieved only in the combined data for one of the detector units (DU1). For the other two detectors, the hypothesis of polarization signal being identical to the one measured from the large region {is excluded at the confidence level of at} least 63\% and 74\% confidence levels (see Table~\ref{tab:brightest}). }

\section{Discussion }
\label{s:discussion}
\begin{figure}
\centering
\includegraphics[angle=0,trim=1.8cm 7.8cm 2.5cm 5cm,clip,width=0.96\columnwidth]{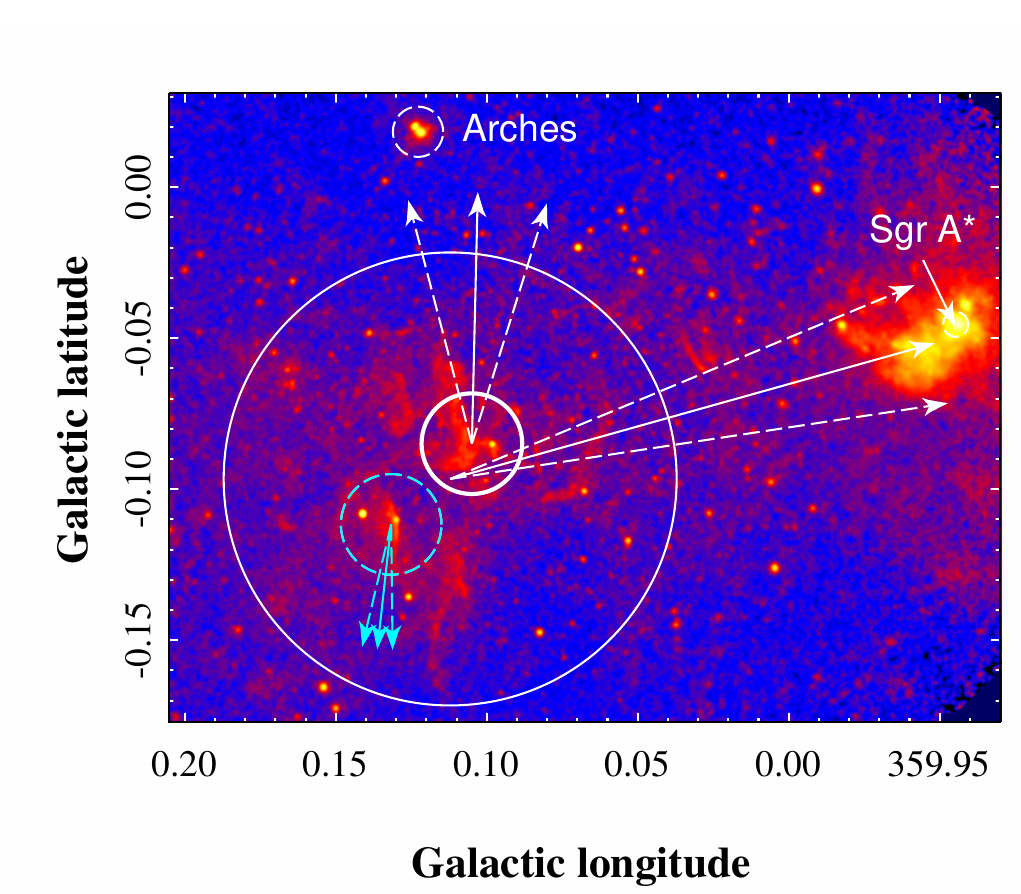}
\caption{\textit{Chandra} (2022-2023)  3--8 keV image (log scale) overlaid with the inferred polarization directions (solid arrows) and with corresponding uncertainties (dashed arrows) converted into either a direction toward the primary source in the reflection case {(large Sgr~A complex and its brightest part in white)} or into the magnetic field direction in the synchrotron scenario for the nonthermal emission from G0.13-0.11 {(in cyan).} The polarization directions, rotated by 180 deg, are identical to the ones shown for the polarization signal, and the signal orientation was chosen to highlight the consistency with the directions toward Sgr~A*, the Arches cluster, or the "wings" of the PWN G0.13-0.11.
}
\label{fig:summ}
\end{figure}

The main driver of the long IXPE observations was to map the polarization signal from individual spatially resolved structures, which are bright in reflected emission. We (marginally) succeeded with this, but at the price of revealing a more sophisticated picture, complicating both data analysis and interpretation. Figure~\ref{fig:summ} presents a summary of the polarization geometry inferred for the three main regions in the Sgr~A direction. 

{
While confirming that the total polarization signal from the large region centred on the Sgr~A complex might come from Sgr~A*, we found a significantly different polarization signal for the brightest and highly time-variable central part of it. We considered two options which, within uncertainties, are consistent with the data.  In one option, the signal from the compact bright region has the same polarization direction as in the large region but a significantly lower PD. In this case, accumulated statistics still allow for significant fluctuations of the PA, even though the probability of this is low. In the second option, the PA is genuinely different from that inferred for the large region.
}

{
The first case leaves open a possibility of Sgr~A* being the primary source of illumination, but it implies that the brightest reflection spot is located at another distance (a lower PD implies larger distance along the line of sight), lending further support to the hypothesis of several flaring episodes \revision{\citep[as already suggested in ][ based on the spatial and temporal variability patterns in different molecular complexes]{2013A&A...558A..32C,2018A&A...610A..34C}}.
}

{
The latter case can be reconciled with a Sgr~A* flare only if its primary emission is strongly polarized with the sky projection of the electric field vector aligned with the source-cloud direction \citep[as shown in the bottom panel of Fig. 2 and in Fig. 4 of ][]{2020MNRAS.498.4379K}. This configuration is possible only when the cloud is located at the far side of the reflecting paraboloid, which allows it to "mirror" the intrinsic polarization of the primary source. Although infrared emission of the regular Sgr~A* flares is known to be polarized \citep[][]{2006A&A...455....1E,2015A&A...576A..20S,2023A&A...677L..10G} and the spatially resolved emission from the currently present accretion flow has been measured \citep[e.g., ][]{2024ApJ...964L..25E}, the possibility of extrapolating these measurements to a hypothetical orders-of-magnitude brighter accretion episode \citep[e.g., ][]{2008MNRAS.391...32D}, which might be responsible for the observed X-ray echo, remains uncertain.}

At the same time, the spectral analysis of the variable diffuse emission from both large and the brightest regions indicates that the spectral shapes of the reflected emission are rather similar, with the slight variations in the equivalent width of the fluorescent line of neutral iron well within the possible effects of different {scattering angle} or metallicity of the reflecting cloud. This appears to disfavor intrinsic polarization of the primary emission since, in this case, apparent rotation of the polarization vector would be associated with substantial variations in the ratio of the fluorescent line to the reflected continuum. In other words, the similarity of the observed spectral shape indicates that the scattering angle in both cases should be similar as well.

Taken at face value, this would indicate the presence of at least two primary sources responsible for the current X-ray echo(es) propagating through the Sgr~A complex. While for the large region Sgr~A* remains a very viable candidate, the polarization vector of the brightest region {is perpendicular to the direction} toward the Arches stellar cluster, which is one of the youngest massive clusters in the Milky Way, and is known to host highly variable compact sources. 

The historical reflection emission from Sgr~B2, Sgr~C, and Sgr~D complexes \citep[][]{2018A&A...612A.102T,2018A&A...610A..34C} is most naturally explained by a flare from a powerful source in the center of the CMZ, i.e., Sgr~A* \citep[e.g.,][]{2013ASSP...34..331P, 2017MNRAS.468..165C}. At the same time, the decay of the reflected emission from these clouds, on a timescale comparable to their light-crossing time, indicates that their illumination was also by a relatively short flare, lasting less than 10 years or so. Also, the spatial distribution of the reflection signal, averaged over 20 yr of available high resolution observations with \textit{Chandra} and \textit{XMM-Newton}, shows that not all molecular complexes in the CMZ were illuminated during this period of time, meaning that the coverage of the CMZ by the illumination front is incomplete, indeed corresponding to a relatively thin volume between two paraboloid surfaces containing only some of the largest molecular complexes \citep[][]{2013ASSP...34..331P,2017MNRAS.468..165C,2018A&A...612A.102T,2022MNRAS.509.6068K,2025A&A...698A.313A}.  
 
Thus, we come to a picture of a plausible "global" illumination of the large molecular complexes in the CMZ by a luminous flare in its centre, and the presence of "local" time--and space-variable reflection signal(s). Although such a picture means a great increase in complexity compared to the simplest "single source, single (short) flare" scenario, it appears to be a possible alternative.

The Galactic Center region is characterized by the highest volume density of all kinds of compact sources, including low-mass and high-mass X-ray binaries and magnetars \citep[e.g.,][]{2020NewAR..8801536S}. Thanks to the decades-long monitoring of this region by several generations of X-ray observatories, we know that many of them are highly variable, and some are transient. The largest concentrations of such objects are expected to be associated with the massive star clusters in this region, namely the Nuclear Star Cluster, Arches, and Quintuplet clusters. While the former hosts a supermassive black hole Sgr~A*, the latter two might host intermediate mass black holes in their centers. A spontaneous accretion event (e.g., due to a partial or total tidal disruption event of a star, planet, or a gas cloud) can, in principle, result in a months-long activation with the luminosity exceeding $10^{40}$ ${\rm erg~s^{-1}}$ \citep[e.g.,][]{2012MNRAS.421.1315Z}. 

Contrary to the case of Sgr~A* being a primary illumination source, it is more difficult to estimate the required fluence of the initial X-ray flare because of the (more) uncertain relative distance between the primary source and the reflector. In particular, for a source located close to the reflecting cloud, the required fluence of the flare will be smaller than the value inferred for Sgr~B2 or Sgr~A complexes in the Sgr~A* scenario. This opens the possibility for even less dramatic transient events to be responsible for the observed `local' reflection echoes.

\revision{For instance, for a cloud located at $Z=+2X$, where $X$ is the projected distance from Sgr A* in the sky plane, i.e., farther away from us than Sgr A*, the degree of polarization will be $\frac{1}{9}$ (see Eq.~\eqref{eq:p}). For the brightest patch, $X\approx 25\,{\rm pc}$.
Similarly to the case discussed above, this scenario requires the scattering cloud located farther away from us than Sgr~A*. In this case, the scattering angle can be close to 180 degrees, and the scattering by itself does not induce any strong polarization. Instead, the reflected emission will have the degree and orientation of polarization corresponding to the incident emission from the primary source.
Among these potential sources of the illuminating X-ray flux, the Quintuplet cluster has the smallest projected distance $X\sim 7 \,{\rm pc}$ to the brightest patch.
This lowers the minimum required luminosity by an order of magnitude compared to the Sgr~A* case (assuming the Quintuplet and the bright patch are at the same distance from us). However, the shorter light crossing time implies that the primary source in the Quintuplet was active some 25 years ago. Adjusting the geometrical parameters, one can shift the flare to $\sim 50$ years ago without a strong increase in the required luminosity. This scenario cannot be unambiguously excluded.}

Further monitoring of the morphological and spectral changes \citep[e.g.,][]{2010ApJ...714..732P,2017MNRAS.465...45C}, in combination with the data on the molecular gas distribution in velocity space \citep[e.g.,][]{2013ASSP...34..331P,2013A&A...558A..32C,2017MNRAS.471.3293C,2020MNRAS.498.4379K,2025ApJ...982L..20B}, should help us to pinpoint the location of the primary source with more confidence, and, as a result, better quantify required energetics of the flare and its possible nature.       

\section{Conclusions}
\label{s:conclusion}

We have reported the results of two deep observations of the currently brightest (in reflected emission) molecular complex Sgr~A taken with the imaging X-ray polarimeter IXPE in 2022 and 2023, and we supplemented them with the data of a quasi-simultaneous \textit{Chandra} monitoring campaign of the same region. We confirm the previous polarization measurement for a large region encompassing the Sgr~A complex with high ($\approx 4\sigma$) significance, but we also revealed an inconsistent polarization pattern for the brightest reflection region in its center. 

{
This compact region has a smaller PD compared to the large region, and the PA is almost perpendicular to the expected direction in the Sgr~A* unpolarized-flare scenario. While the statistical significance of this result is at the level of less than 3$\sigma$, it might indicate that several illumination fronts have propagated through the CMZ simultaneously, with at least one of them not being due to a Sgr~A* flare. The scenario of Sgr~A* as the primary source of illumination for both regions is still possible, but it requires either (i) that the currently brightest compact region of the Sgr~A complex is located at a significantly farther distance (and hence is illuminated by an older flare) or (ii) that the primary emission was itself strongly polarized in a manner allowing almost perpendicular rotation of the polarization vector relative to the unpolarized case \citep[e.g., in the configuration illustrated in ][]{2020MNRAS.498.4379K}. In fact, the latter option also requires the reflecting medium to be well "behind"  Sgr~A*.
}

{
If instead the inferred polarization direction is indeed determined by the direction to the primary source, it could be associated with the Arches stellar cluster or a currently unknown transient source located in a closer vicinity to the illuminated cloud \citep[see a discussion of implications of such a scenario in ][]{1998MNRAS.297.1279S}. Although a scenario of multiple (transient) illuminating sources is straightforward to model \citep[e.g.,][]{2014A&A...564A.107M}, the arising complexity of spectral variability and polarization patterns requires sensitive monitoring observations of many individual reflecting complexes \citep[e.g.,][]{2014MNRAS.441.3170M}. }

{
Significantly deeper observations with IXPE would be required to unequivocally distinguish between the scenarios. Given the consistency of the observed noise levels with the expected statistical fluctuations, accumulation of an additional $\approx$1.1 Ms of clean exposure time would be needed to ensure the $3\sigma$ confidence level detection for the brightest reflection region (if the real signal is indeed at the $2.4\sigma$ detection level \revision{and would stay constant in time). There is, however, an indication of a decrease in the observed surface brightness from the Sgr~A complex, which makes this task even more challenging.} 

A combination of continued high-resolution imaging, micro-calorimetric spectroscopy \citep[e.g., with XRISM observatory,][]{2025PASJ..tmp...28T}, and global monitoring of the new reflection regions via CMZ-wide surveys \citep[e.g.,][]{2020MNRAS.498.4379K,2024MNRAS.529..941S} offers a promising path forward \citep[along with the roadmap outlined in ][]{2019BAAS...51c.325C}. Next-generation X-ray polarimetry missions will be required to disentangle the full complexity of X-ray reflection signals if the scenario of multiple sources and/or multiple flares is confirmed. An additional window might be offered by measuring polarization and spectra at higher energies (>10 keV), where the reflected continuum should be dominant over other emission components. Prospective missions such as the Enhanced X-ray Polarimetry Observatory, an evolution of the NGXP concept \citep[][]{2021ExA....51.1109S}, would be ideally suited for this task.}

\vspace{0.1cm}

\begin{acknowledgements}
The Imaging X-ray Polarimetry Explorer (IXPE) is a joint US and Italian mission.  The US contribution is supported by the National Aeronautics and Space Administration (NASA) and led and managed by its Marshall Space Flight Center (MSFC), with industry partner Ball Aerospace (contract NNM15AA18C).  The Italian contribution is supported by the Italian Space Agency (Agenzia Spaziale Italiana, ASI) through contract ASI-OHBI-2022-13-I.0, agreements ASI-INAF-2022-19-HH.0 and ASI-INFN-2017.13-H0, and its Space Science Data Center (SSDC) with agreements ASI-INAF-2022-14-HH.0 and ASI-INFN 2021-43-HH.0, and by the Istituto Nazionale di Astrofisica (INAF) and the Istituto Nazionale di Fisica Nucleare (INFN) in Italy.  This research used data products provided by the IXPE Team (MSFC, SSDC, INAF, and INFN) and distributed with additional software tools by the High-Energy Astrophysics Science Archive Research Center (HEASARC), at NASA Goddard Space Flight Center (GSFC).

{We thank Dylan Maurel of the IXPE Science Operations Center for manually reprocessing the IXPE data.}

IK acknowledges support by the COMPLEX project from the European Research Council (ERC) under the European Union’s Horizon 2020 research and innovation program grant agreement ERC-2019-AdG 882679.
JS thanks GACR project 21-06825X. AV, RK, and WF acknowledge support from the Smithsonian Institution and the Chandra High Resolution Camera Project through NASA contract NAS8-03060. 
IL was funded by the European Union ERC-2022-STG -- BOOTES -- 101076343.
RF, ECo, ADM, PSo, FLM, FMu, and SF are partially supported by MAECI with grant CN24GR08 “GRBAXP: Guangxi-Rome Bilateral Agreement for X-ray Polarimetry in Astrophysics”.

We are grateful to the anonymous referee for a constructive report that helped us to improve the paper.
\end{acknowledgements}


\bibliographystyle{aa}
\bibliography{main.bib} 



\begin{appendix}

\section{Regions of interest}
\label{s:regions}
\revision{
Here we present a table that summarizes the parameters of the regions of interest used for the signal and (astrophysical) background estimations throughout the paper. This refers to the light curves in Sect.~\ref{s:imaging} and the spectral and polarization analysis in Sect.\ref{s:polarimetry}}.

\begin{table}[]
\centering
\caption{\revision{Parameters of circular regions of interest, including their short names, equatorial (RA, Dec) and Galactic ($l,b$) coordinates of their centers, and radii (in arcsec). }}
\begin{tabular}{lllllr}
\hline\hline
  \multicolumn{1}{c}{Region} &
  \multicolumn{1}{c}{RA,$^\circ$} &
  \multicolumn{1}{c}{Dec,$^\circ$} &
  \multicolumn{1}{c}{$l$,$^\circ$} &
  \multicolumn{1}{c}{$b$,$^\circ$} &
  \multicolumn{1}{c}{$r,``$} \\
\hline
Arches & 266.460 & $-$28.822 & 0.123 & 0.018 & 30\\
Background & 266.628 & $-$28.801 & 0.217 & $-$0.097 & 90\\
G0.13-0.11~(nt) & 266.592 & $-$28.882 & 0.132 & $-$0.112 & 60\\
Sgr~A & 266.566 & $-$28.891 & 0.112 & $-$0.097 & 270\\
Sgr~A~(re) & 266.550 & $-$28.891 & 0.105 & $-$0.085 & 60\\
Sgr~A~(bg)  & 266.579 & $-$28.849 & 0.154 & $-$0.085 & 60\\
\hline\end{tabular}
\tablefoot{In most of the analysis procedures, regions G0.13-0.11~(nt) and Sgr~A~(re) are excluded from a bigger Sgr~A region encompassing them unless the opposite is explicitly stated.}
\label{t:regions}
\end{table}
\section{Systematic uncertainties}
\label{s:systematic}

\begin{figure*}
\centering
\includegraphics[angle=0,trim=1.cm 7.cm 1cm 9.cm,clip,width=2.\columnwidth]{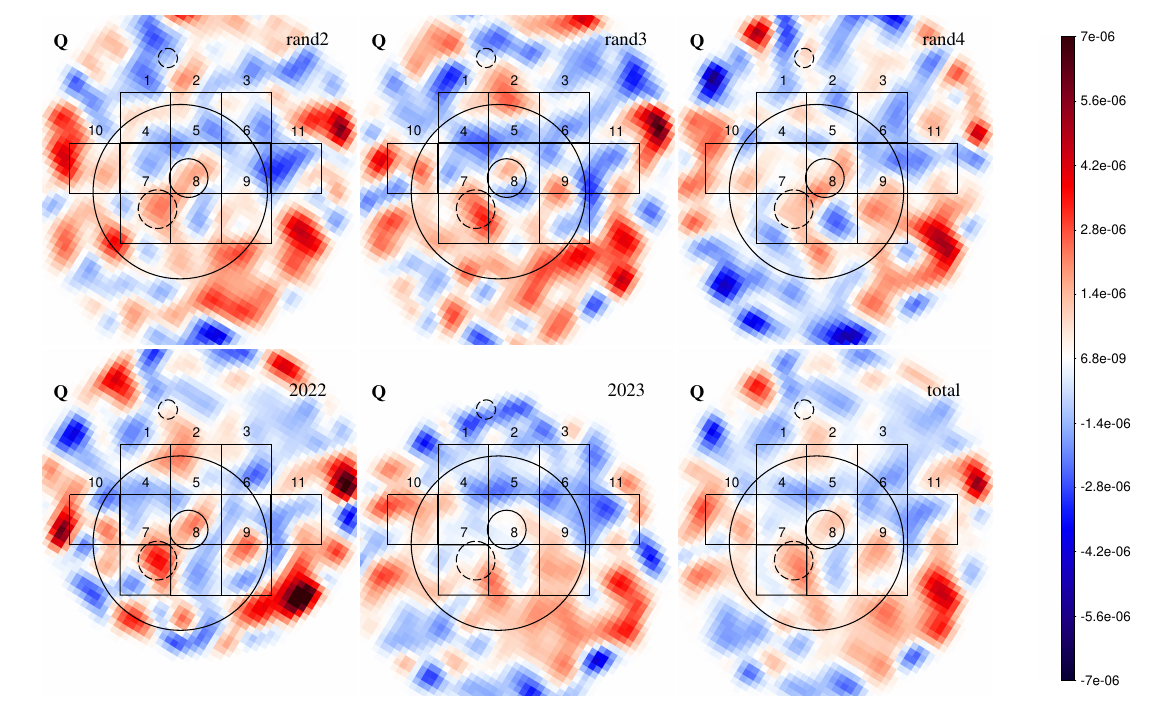}
\caption{IXPE 3--6 keV Stokes $Q$ images (${\rm ct~s^{-1}~pixel^{-1}}$, linear scale, Galactic coordinates) for various subsets ($\sim$half) of the data (3 random halves, observations in 2022 and 2023) compared to the total dataset. Eleven square regions are used to estimate statistical variance at a scale of comparable size to the extraction regions for the brightest reflection spot. The large circle is 4.5$\arcmin$ radius, and the small solid circles mark positions of G0.13-0.11 and the brightest reflection region. All images are on the identical linear scale with white color corresponding to zero.
}
\label{fig:ixpe_jack_q}
\end{figure*}
\begin{figure*}
\centering
\includegraphics[angle=0,trim=1.cm 7.cm 1cm 9.cm,clip,width=2.\columnwidth]{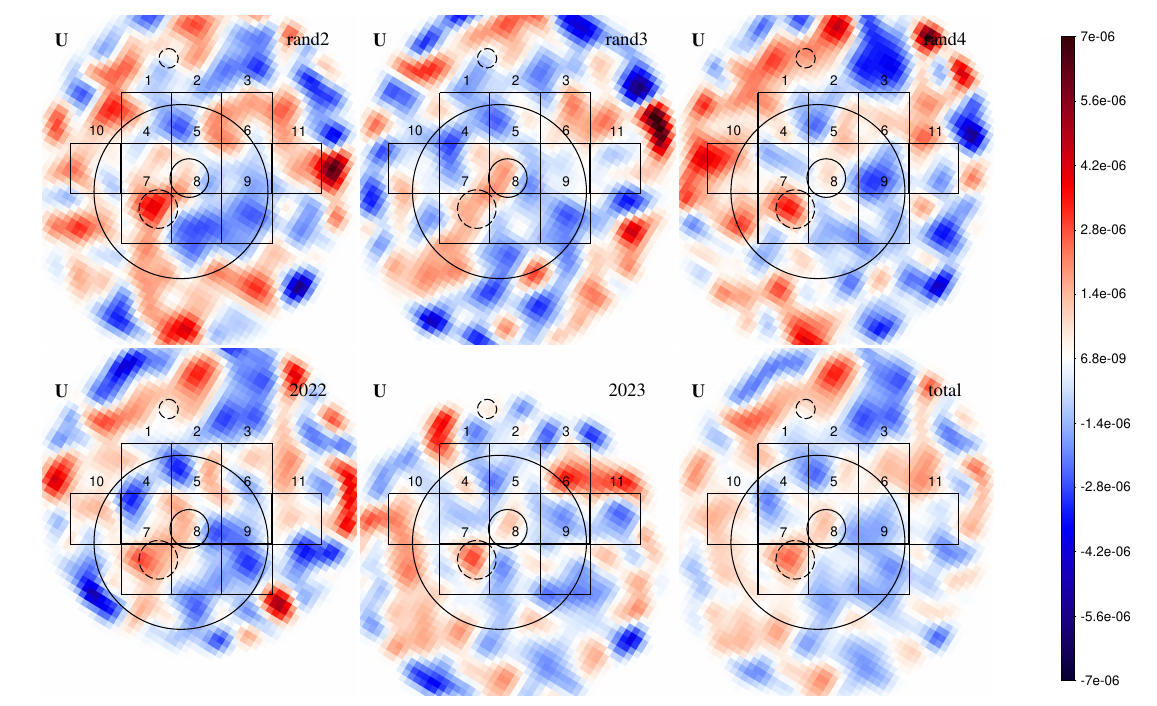}
\caption{Same as Fig.~\ref{fig:ixpe_jack_q} but for Stokes $U$.
}
\label{fig:ixpe_jack_u}
\end{figure*}

The IXPE observations of the Sgr~A complex are aimed at faint, spatially and temporally variable diffuse emission in the vicinity of a number of bright point-like and extended sources. As a result, measurement of the polarization signal and especially mapping of its spatial morphology is very challenging and sensitive to uncertainties in the current knowledge of systematic effects inherent to such observations.

Here, we estimate the level of systematic uncertainties that might be present in the data in addition to the purely statistical noise caused by limited count number statistics. In order to do so, we perform a number of jack-knifing experiments by considering variations in the obtained results if various subsets of the full IXPE dataset are included. In particular, one can compare data collected in two epochs separately, by separate Detector Units (DU) of IXPE, as well as spatial variations in excess of the expected statistical noise level. We calculate variance within regions of relevant size (comparable to the brightest in the reflection region) shown in Figs.~\ref{fig:ixpe_jack_q} and \ref{fig:ixpe_jack_u} for Stokes $Q$ and $U$, respectively.

The signal-to-noise ratios obtained by dividing the extracted average Stokes values from the regions of interest (the brightest reflection spot and G0.13-0.11 region for reference) by the estimated variances from the maps are consistent with the ones obtained from spectral fitting. None of the observation epochs appears to be different from the full dataset in excess of the expected variance, which is estimated by analyzing variances between different halves of the data selected randomly. Thus, we conclude that statistical properties of the Stokes $Q$ and $U$ maps are consistent with expected uncorrelated noise behavior (at least on the spatial scale considered here), meaning that background and signal fluctuations dominate over possible systematic effects.

In order to quantify possible systematic effects of the observational setup or detectors, we also quantify a variance in the polarization signal inferred from the data of individual observation epochs and detector units. Each data subset contains $\approx6=2{\rm(epochs)\times3(DU)}$ times smaller effective exposure time, boosting the statistical noise in each individual measurement, especially for the Stokes $Q$ and $U$ maps. No significant excess variation of any detector signal in any epoch is observed in this test as well.

\begin{table*}[]
\centering
\caption{A summary of polarimetric results for the large $r=$4.5$\arcmin$ region encompassing Sgr~A complex (excluding $r=$1$\arcmin$ regions around G0.13-0.11 and the brightest reflection spot)
in the 3 -- 6 keV energy band. 
}
\label{tab:all_minus_filament_and_brightest}
\begin{tabular}{ccccccc}
\hline
{Dataset}      & {$q$, \%}     & {$u$, \%}     & {PD, \%}   & {PA, °}    & {$\Sigma$}   & {CL, \%}\\
\hline
{Obs 1}        &$-$1.0±1.4          &$-$5.7±1.4          & 6.0±1.4          &$-$50±7                       & 4.0                 & 99.97         \\
{Obs 2}        & 1.8±1.4           &$-$3.4±1.4          & <7.5   & NC                 & 2.7                           & 97.47                  \\
{Combined}     & 0.4±1.0           &$-$4.5±1.0          & 4.6±1.0          &$-$42.5±6                         & 4.5                 & 99.99          \\
{Combined DU1} & 1.1±1.7           &$-$3.9±1.7          & <8.5   & NC                             & 2.4                 & 93.97                  \\
{Combined DU2} & 1.3±1.8           &$-$6.2±1.8          & 6.0±2.0          &$-$39±8                       & 3.6                 & 99.82         \\
{Combined DU3} &$-$1.3±1.8          &$-$3.5±1.8          & <8.0   & NC                              & 2.2                 & 90.23                   \\
\hline
\end{tabular}
\tablefoot{The results are split for the first and second observations separately, their combination, and the combined results for each IXPE Detector Unit (DU).
The normalized Stokes $q$=Q/I and $u=Q/I$ parameters, and the corresponding PD, are determined with respect to the total intensity in the 3--6 keV band. It is different from the PD of the reflected continuum only and is intended to demonstrate consistency and variations between individual datasets.
Whenever {the detection's statistical significance level $\Sigma$} is below the 3$\sigma$ threshold (unaffected by the choice of relative normalization), the 1$\sigma$ upper limit on the PD is provided, whereas the PA is non-constrained (NC).}
\end{table*}
\begin{table*}[]
\centering
\caption{Same as Table \ref{tab:all_minus_filament_and_brightest} but for the brightest reflection region. { }}
\label{tab:brightest}
\begin{tabular}{ccccccccc}
\hline
{Dataset}      & {$q$, \%} & {$u$, \%} & {PD, \%}  & {PA,°} & { $\Sigma$} & {CL, \%}& {CL$_0$, \%} \\
\hline
{Obs 1}        & 4.3±4.9       & 8.2±4.9       & <22 & NC            & 1.9        & 83.06   &    98.59         \\
{Obs 2}        & $-$2.8±5.1     & 4.9±5.1       & <19 & NC            & 1.1        & 45.55    &   79.91          \\
{Combined}     & 0.9±3.5       & 6.6±3.5       & <16 & NC            & 1.9        & 83.18     &   99.05         \\
{Combined DU1} & $-$10.9±6.0     & 16.5±6.0      & 20±6        & 62±9      & 3.3        & 99.71   & 99.93   \\
{Combined DU2} & 6.2±6.2       & 1.5±6.2       & <22 & NC            & 1          & 41.35      & 63.19 \\
{Combined DU3} & 8.1±6.2       & 1.2±6.2       & <24 & NC            & 1.3        & 58.59   & 73.42          \\
\hline
\end{tabular}
\tablefoot{An additional column, CL$_0$, shows confidence level at which null hypothesis of the data being consistent the model inferred or the large $r=$4.5$\arcmin$ region encompassing Sgr~A complex (excluding $r=$1$\arcmin$ regions around G0.13-0.11 and the brightest reflection spot) can be ruled out (assuming similar contribution of the reflected emission).}
\end{table*}

\begin{table*}[]
\centering
\caption{\revision{Similar to Table~\ref{tab:all_minus_filament_and_brightest} for the large $r=$4.5$\arcmin$ region encompassing Sgr~A complex excluding only $r=$1$\arcmin$ region around G0.13-0.11 and aimed at comparison of the different observational epochs.}
}
\label{tab:all_minus_filament}
\begin{tabular}{ccccccc}
\hline
{Dataset}      & {$q$, \%}     & {$u$, \%}     & {PD, \%}   & {PA, °}    & {$\Sigma$}   & {CL, \%}\\
\hline
{Obs 1}        &$-$0.6±1.4          &$-$4.6±1.4          & 4.6±1.4          &$-$50±7                       & 3.4                & 99.6         \\
{Obs 2}        & 1.5±1.4           &$-$2.8±1.4          & 3.2±1.4   &  $-$42±7.5                & 2.3                           & 93.05                  \\
{Combined}     & 0.4±1.0           &$-$3.7±1.0          & 3.7±1.0          &$-$42±7.5                         & 3.8                 & 99.93          \\
\hline
\end{tabular}
\end{table*}


\end{appendix}
\label{lastpage}
\end{document}